\documentclass[prd,preprint,superscriptaddress,nofootinbib,showpacs]{revtex4-2} 
\usepackage{graphicx,amsmath,amsfonts,amssymb,revsymb,dcolumn,relsize,hyperref}

\usepackage{color}
\hypersetup{colorlinks=true,citecolor=blue,urlcolor=blue,linkcolor=blue}

\newcommand{\be}{\begin{equation}}
\newcommand{\ee}{\end{equation}}
\newcommand{\bea}{\begin{eqnarray}}
\newcommand{\eea}{\end{eqnarray}}

\begin{document}

\global\long\def\imsize{0.83\columnwidth}
 \global\long\def\halfsize{0.45\columnwidth}
\title{Probing spatial orientability of Friedmann--Robertson--Walker spatially flat spacetime}

\author{N.A. \surname{Lemos}} \email{nivaldolemos@id.uff.br}
\affiliation{Instituto de F\'{\i}sica, Universidade Federal Fluminense,
Av. Litor\^anea, S/N \\
24210-340 Niter\'oi -- RJ, Brazil}
\author{D. \surname{M\"uller}} \email{dmuller@fis.unb.br}
\affiliation{
Instituto de F\'{\i}sica, Universidade de Bras\'{\i}lia
70919-970 Bras\'ilia - DF,  Brazil
}
\author{M.J. \surname{Rebou\c{c}as}} \email{reboucas.marcelo@gmail.com}
\affiliation{Centro Brasileiro de Pesquisas F\'{\i}sicas,
Rua Dr.\ Xavier Sigaud 150 \\
22290-180 Rio de Janeiro -- RJ, Brazil}

\renewcommand{\baselinestretch}{0.96}

\date{\today} 

\begin{abstract}
One important global topological property of a spacetime manifold  is orientability.
It is widely believed that spatial orientability can only be tested by global journeys
around the Universe to check for orientation-reversing closed paths.
Since such global journeys are not feasible, theoretical arguments that combine
universality of physical experiments with local arrow of time, CP violation and CPT
invariance are usually offered to support the choosing  of time- and space-orientable
spacetime manifolds.  The nonexistence of globally defined spinor fields on
a non-orientable spacetime is another theoretical argument for orientability.
However, it is conceivable that orientability can be put to test by local
physical effects.
In this paper, we show that it is possible to locally access spatial orientability of
a spatially flat Friedmann--Robertson-Walker spacetime through
quantum vacuum electromagnestic fluctuations. We argue that a putative
non-orientability of the spatial sections of spatially flat  FRW spacetime can be
ascertained by the study of the stochastic motions of a charged particle or a
point electric dipole under quantum vacuum electromagnetic fluctuations.
In particular, the stochastic motions of a dipole permit the recognition
of a presumed non-orientability of $3-$space in itself.
\end{abstract}

\pacs{03.70.+k, 05.40.Jc, 42.50.Lc, 04.20.Gz, 98.80.Jk,  98.80.Cq}
\maketitle
\newpage
\section{Introduction} \label{Intro}
The standard  approach to model the Universe starts with two  
basic assumptions. First, Weyl's principle~\cite{Weyl:1923yve,weyl2009republication} is postulated, which entails\footnote{We adopt here a formulation of  Weyl's principle in which it is assumed 
that world lines of \textit{fundamental particles} (galaxies, galaxy clusters)
form, on average, a congruence  of non-intersecting diverging geodesics  
emerging from the distant past and orthogonal to
space-like hypersurfaces $M_3$.
In this form  the principle permits a comoving frame relative to which the
constituents of the universe are at rest
on average~\cite{Robertson:1933zz,Narlikar:2002wxx,Rugh:2010mv}.}
the existence of a cosmic 
time $t$. Second, it is assumed that our 
$3-$dimensional space is homogeneous and isotropic (cosmological principle).
The most general spacetime geometry that embodies  these assumptions is
the Friedmann--Robertson--Walker (FRW) metric
\begin{equation} \label{RWmetric}
 ds^2 = c^2 dt^2 - a^2 (t) \left[\, \frac{dr^2}{1-kr^2} +
r^2(d\theta^2 + \sin^2 \theta  d\phi^2) \,\right]\,,
\end{equation}
where $c$ is the speed of light, $a(t)$ is the  scale factor, and
the spatial curvature is specified by the constant $k$, which takes the
values $k=0,\pm 1$ for Euclidean, spherical and hyperbolic geometries,
respectively.

The metric~(\ref{RWmetric}) expresses locally the two above basic assumptions.
It does not  specify the topology of the spacetime manifold $\mathcal{M}_4$
or of the corresponding spatial ($t= \mbox{const}$) section $M_3$. 
However, the FRW metric~\eqref{RWmetric} is consistent with the global decomposition
$\mathcal{M}_4 = \mathbb{R} \times M_3$, which we assume in this work.

Regarding the spatial geometry, recent high precision cosmic microwave background
radiation (CMB) data from the Planck satellite~\cite{2016,2020}
have provided strong evidence that the universe
is very nearly flat with curvature parameter $|\Omega_k| < 0.003$,
which is compatible with the standard inflationary predictions
that the spatial curvature should be very small today.  
These indications 
support the assumption we  make in this work that the spatial section $M_3$
is flat ($k=0$).

As to the topology of the spatial sections, we first note that the FRW
geometry~\eqref{RWmetric} constrains but does not specify the topology of $M_3$.
In this way, no classical geometric theory as, for example, general relativity can be
used to derive the $M_3$ topology.  
However, for $k=0$, it is a  mathematical fact  that, in addition to the simply-connected
Euclidean space $\mathbb{E}^{3}$, there are $17$ topologically inequivalent quotient flat
manifolds  with nontrivial topology~\cite{wolf1967spaces,Thurston+2014}.
Given this  set of topological possibilities for $M_3$  
and despite our present-day inability to infer the topology from a
fundamental theory, as for example quantum gravity~\cite{1987IAUS..124..461F},
to disclose the spatial topology of FRW spacetime we must rely ultimately on
cosmological observations (see the review articles~%
\cite{Ellis:1970ey,LACHIEZEREY1995135,Starkman:1998qx,LEVIN2002251,%
Reboucas:2004dv,Luminet:2016bqv}) or  on local physical
experiments.%
\footnote{
For recent constraints on cosmic topology from CMBR data  we refer the readers to
Refs.~\cite{2014,2016,Cornish:2003db,ShapiroKey:2006hm, Bielewicz:2010bh,Vaudrevange:%
2012da,Aurich:2013fwa}
For some limits on the circles-in-the-sky method designed for the searches of spatial
topology through CMBR see Ref.~\cite{Gomero:2016lzd}}

Topological properties precede the geometrical features of a manifold.
Thus, it is important to find out whether, how and to what extent physical
results depend on a nontrivial topology.

Nonstandard choices of the background spatial topology
affect the mean squared velocity of  charged test particles
under quantum vacuum fluctuations of the electromagnetic field.
In fact, on the   assumption that the net role played by the spatial topology
is more clearly ascertained in the static FRW flat spacetime, the question of
how a nontrivial topology of the spatial section of Minkowski spacetime
modifies the stochastic motions of a test charged particle and a point
electric dipole under quantum vacuum
fluctuations of the electromagnetic field was studied in  recent
papers~\cite{Bessa:2019aar,Lemos:2020ogj}.
By the way, we mention that the case of a point particle coupled to
a massless field living in a topologically nontrivial space
was considered in Ref.~\cite{Matas:2015rba}.

Orientability is an important topological property of spacetime manifolds.
It is generally assumed that, being a global property, the orientability of
$3-$space cannot be tested locally. So, a test for spatial orientability
would require a global trip along some specific closed paths around the
whole $3-$space to check whether one returns with left- and right-hand
sides exchanged.
Since such a global expedition does not seem to be feasible at the
cosmological scale, theoretical arguments that combine universality of
physical experiments with local arrow of time, CP violation and CPT invariance
are usually invoked~\cite{1967JETPL...6..236Z,Hawking:1973uf,%
penrose1984spinors,1979grec.conf..212G} to support the choosing of time-
and space-orientable manifolds, although there are dissenting stances%
~\cite{Hadley:2002fr,HadleyM2018}.
The impossibility of having \textsl{globally} defined spinor fields
on non-orientable spacetime  manifolds \cite{Geroch:1968zm,Geroch:1970uv}.
is another theoretical argument to support the choice of space-and-time
orientable manifolds.%
\footnote{One can certainly take advantage of  theoretical arguments of this sort to
support such underlying assumptions, but not as a replacement to experimental and
observational evidence in physics.}

Since  $8$ out of the $17$  possible  flat $3-$manifolds, $M_3$, with
nontrivial topology are non-orientable~\cite{wolf1967spaces}, the question as to
whether velocity fluctuations could be also employed to locally reveal  specific
topological properties such as orientability was examined in
Ref.~\cite{Lemos:2020ogj}.
It was shown that it is possible to \textsl{locally} access the spatial
orientability of Minkowski spacetime through the study of the stochastic motions
of a charged particle and a point electric dipole subject to these fluctuations
in Minkowski spacetime with orientable and  non-orientable spatial topologies.
It was found that a characteristic inversion pattern exhibited by certain
 statistical orientability indicator curves, constructed from the mean
square velocity of an electric dipole, can be used as a  local physical
signature of non-orientability of the spatial section $M_3$ of Minkowski
spacetime.    

Thus, a question that naturally arises is how these results are
modified in an expanding FRW universe whose curvature parameter is
within the bounds determined by Planck data~\cite{2014,2016},
which indicate that a flat geometry is a good approximation to model the
spatial section of the Universe in the framework of general relativity.
To tackle this question,
in this paper we study the stochastic motions of a charged particle and a
point electric dipole under quantum vacuum electromagnetic fluctuations
in a spatially-flat FRW  geometry with spatial sections  endowed with
an orientable and its counterpart non-orientable spatial topologies.
In so doing we extend the results of~\cite{Bessa:2019aar,Lemos:2020ogj}
from the static Minkowski spacetime to a dynamical FRW spacetime. Our result for the orientability
indicator is very general, in the sense that it does not depend on the underlying gravitational theory,
because it is  obtained from first topological-geometrical principles alone.

The structure of the paper is as follows.
In Sections~\ref{TopSet} and~\ref{Syst-montion} we describe the topological
and dynamical settings, respectively. In Sections~\ref{disp-particleE17}
and~\ref{Dipole-motion} we derive
statistical
orientability indicators both for the charged particle and the point
electric dipole that are independent of any specific metrical
theory of gravity.
To concretely study the time evolution of the orientability indicators,
in Section~\ref{casestudy} we choose general relativity and a barotropic
perfect fluid as the matter content.
In the case of a charged particle, we show that it is possible to distinguish
the orientable from the non-orientable topology by comparing the time
evolution of an orientability indicator, defined from  the stochastic motion
of the particle, in the orientable and non-orientable topologies.

We then turn to the more substantial problem of finding a way to decide about
the orientability of a given spatial manifold in  itself, without having to
compare the results for a non-orientable space with those for its orientable
counterpart.
Motivated by a dipole's directional properties, we inquire  whether the
stochastic motions of a point electric dipole would be more effective
to unveil the presumed non-orientability of a $3-$space in itself.
From the orientability indicators  computed for the dipole we identify
a characteristic inversion pattern displayed by the  orientability
indicator curves for the non-orientable topology,
implying that the putative non-orientability  can be detected
per se. 

In Section~\ref{Finals} we present our final remarks and summarize
our findings, which indicate that it may be possible to locally disclose
a conceivable spatial non-orientability of FRW spacetime
through the stochastic motions of point-like ``charged'' objects
under  quantum vacuum fluctuations of the electromagnetic field.

\section{Topological fundamentals }  \label{TopSet}

To make this work to a certain extent self-contained, in this section
we define the notation, give some basic definitions and present a few
results concerning the topology of flat three-dimensional manifolds
which are used in this paper.

We begin by recalling that in the standard cosmological model
the spacetime is a manifold $\mathcal{M}_4$ locally endowed
with a FRW metric~\eqref{RWmetric} and globally decomposable
as  $\mathcal{M}_4 = \mathbb{R}\times M_3$.  
Although the spatial section $M_3$, whose geometry we assume to be Euclidean,
is usually taken to be the simply-connected Euclidean space $\mathbb{E}^{3}$,
it can likewise be one of the possible $17$ topologically inequivalent
multiply-connected  quotient manifolds $\mathbb{E}^3/\Gamma$ where $\Gamma$
is a  discrete group of 
isometries or holonomies acting freely on the covering manifold
$\mathbb{E}^{3}$~\cite{wolf1967spaces,Thurston+2014}.
The quotient manifolds are compact in at least one direction.
The action of $\Gamma$ tiles the noncompact covering space $\mathbb{E}^{3}$
into infinitely many identical copies of the  fundamental domain (FD) or cell (FC).
Thus, the multiple connectedness of the quotient manifold gives rise to periodic
boundary conditions (repeated domains or cells) on the covering manifold
$\mathbb{E}^{3}$  that are determined by the action of the group $\Gamma$
on $\mathbb{E}^{3}$.

An example of flat quotient manifold is the so-called slab space,
denoted in the literature by $E_{16}$, which is open (noncompact) in two
independent directions and decomposed into
$E_{16} = \mathbb{R}^2  \times \,\mathbb{S}^1  =\mathbb{E}^3/\Gamma$,
where  $\mathbb{R}^2$ and $\mathbb{S}^1$ stand for the real plane and
the circle, respectively.
A fundamental domain is
a slab  with a pair of opposite faces (two infinite parallel planes)
identified through translations.
The simply-connected covering space $\mathbb{E}^3$ is tiled with
these equidistant parallel planes, which together with two
noncompact independent spatial directions form the FD of $E_{16}$.
The periodicity in the  compact direction is 
given by the circle $\mathbb{S}^1$, whereas the noncompact
independent directions form $\mathbb{R}^2$.

In forming the quotient manifolds $M_3$ an essential  point is that  
they are obtained from the covering manifold $\mathbb{E}^3$ through identification
of points that are equivalent under the action of the group $\Gamma$. In this way,
each point in the quotient manifold $M_3$ represents all the equivalent points in
the covering space. Thus, for example, for $E_{16}$ quotient space,
taking the $x$-direction as compact,  one has that, for
$n_x \in \mathbb{Z}$ and compact length  $L>0$, points $(x,y,z)$ and
$(x+n_xL, y, z)$ are identified. In terms of the covering isometry
$\gamma \in \Gamma$ one has
\begin{equation}
\label{ident-E16}
P = (x,y,z) \,\mapsto \, P^{\prime} = \gamma P = (x + n_x L,\, y, z )\,.
\end{equation}
Another example that we shall be concerned with in this paper is
the slab space with flip $E_{17}$, which involves an additional
inversion in a direction orthogonal to the compact direction, that is,
one direction in the tiling planes is flipped as one moves from
one plane to the next. Taking the $x$-direction as compact
and letting the flip be in the $y$-direction, in the covering space
$\mathbb{E}^3$ one has
\begin{equation}\label{ident-E17}
P = (x,y,z)  \; \mapsto \;  P^{\prime} = \gamma P = (x+n_xL,\,(-1)^{n_x}y,\,z),
\end{equation}
and the identification $P \equiv P^{\prime}$ defines the $E_{17}$ topology.

It should be noted that for each of the $17$ quotient manifolds,
$\mathbb{E}^3/\Gamma$, the associated  periodic conditions on the covering
space $\mathbb{E}^3$ are determined by the group $\Gamma$, and clearly different
discrete isometry groups $\Gamma$ define different topologies for $M_3$, which in
turn give rise to different periodicities  and associated tiling of the
covering space  $\mathbb{E}^3$.

Unlike the local geometric concept of homogeneity, which is formulated in
terms of the action of the local group of isometries, in topological spaces
we have the concept of  \textsl{global} or \textsl{topological homogeneity}.
A way to describe global topological homogeneity  of the   
quotient manifolds is through distance functions. In fact, for any
$\mathbf{x} \in M_3$ the distance function $\ell_\gamma (\mathbf{x})$
for a given isometry $\gamma \in \Gamma$ is def\/ined by  
\be
\label{dist-function}
\ell_\gamma(\mathbf{x}) = d(\mathbf{x}, \gamma \mathbf{x}) \; ,
\ee
where $d$ is the Euclidean metric.  The distance function   
gives the length of the closed geodesic that passes through $\mathbf{x}$
and is associated with a holonomy $\gamma \in \Gamma$.
For a  \textsl{globally homogeneous} manifold, endowed with a topology defined
by a group $\Gamma$, the distance function for any covering
isometry $\gamma \in \Gamma$ is constant. In globally inhomogeneous manifolds, i.e.
manifolds with inhomogeneous topologies,  in contrast,
the length of the closed geodesic associated with at least one  $\gamma$
is non-translational (screw motion or flip, for example) and the corresponding
distance depends on the point $\mathbf{x} \in M_3$, and then is not constant.
In this way, the slab space $E_{16}$   is globally homogeneous since all
$\gamma$s are translations, whereas the slab space with  flip, $E_{17}$,
is  globally inhomogeneous since the covering group $\Gamma$ contains a
flip, which clearly is a non-translational holonomy.

Another very important global (topological) property of manifolds that
we shall deal with in this paper is \textsl{orientability}, which measures whether
one can consistently choose  a definite orientation for loops in a manifold.
An orientation-reversing path in a manifold $M_3$ is a path that brings a
traveler back to the starting point mirror-reversed. Manifolds that contain
an orientation-reversing path are \textsl{non-orientable}, whereas those
that do not have any such reversing path are called   
\textsl{orientable}~\cite{weeks2020shape}.
In two dimensions, one has planes, cylinders and two-tori as examples
of orientable surfaces, whereas the M\"obius strip and Klein bottle are
non-orientable surfaces.
For three-dimensional quotient manifolds, when the covering group $\Gamma$
contains at least one holonomy $\gamma$ that is a reflection (flip) the
associated quotient manifold is non-orientable.
In this way, the slab space, $E_{16}$,  is orientable while the slab space with
flip, $E_{17}$, is non-orientable. Clearly  non-orientable manifolds are necessarily
topologically inhomogeneous as the covering group $\Gamma$ contains
a reflexion, which is a non-translational covering holonomy.

In Table~\ref{Tb-2-Orient_and_Non_orient} we collect the names and symbols
 used to refer to the manifolds together with the number of compact
independent dimensions and information concerning their orientability
and global homogeneity.
In the next section we shall study the motions of a charged test particle
and a point electric dipole under quantum vacuum fluctuations of the
electromagnetic field in the expanding FRW spacetime whose spatial
sections are the manifolds given in Table~\ref{Tb-2-Orient_and_Non_orient}.
\begin{table}[h!]
\begin{tabular}{lcccc} 
\hline\hline
 Name   &   Symbol             & $\quad$ Compact Dim.\ & $\quad$ Orientable  &$\quad$ Homogeneous \\
\hline
Slab space           &       $E_{16}$    &   $1$       &     yes                &  yes \\
Slab space with flip &  $E_{17}$         &   $1$       &     no                 &  no \\
\hline
\end{tabular}
\caption{Names and symbols of two Euclidean orientable and 
non-orientable  quotient manifolds $M_3 =\mathbb{E}^3/\Gamma$ together with the number of compact
dimensions (Comp.), orientability and global (topological) homogeneity.}
\label{Tb-2-Orient_and_Non_orient}    
\end{table}

Finally, we briefly mention a few results that are 
used  throughout this paper (for a detailed discussion we refer  the reader
 to Ref.~\cite{1979grec.conf..212G}).
All simply-connected spacetime manifolds are both time- and space-orientable.
The product of two manifolds is simply-connected if and only if
the factors are.
If the spacetime is of the form $\mathcal{M}_4 = \mathbb{R}\times M_3$ then
space-orientability of the spacetime reduces to  orientability of the $3-$space
$M_3$. This applies to the spacetime endowed both with the  $E_{16}$ (orientable)
and the $E_{17}$ (non-orientable) topology that we deal with in this paper.

\section{Non-orientability from electromagnetic fluctuations}
\label{Syst-montion}

As shown in \cite{Bessa:2019aar,Lemos:2020ogj}, nontrivial spatial
topologies influence the stochastic motions both of a charged particle and a point
electric dipole  in the presence of quantum vacuum fluctuations of the electromagnetic
field in Minkowski spacetime.
Here we investigate how these results are modified if instead of  static Minkowski
spacetime an expanding FRW flat universe is the background geometry for the motions of
the charged particle and the dipole. To this end we consider a spatially flat FRW
spacetime endowed with two inequivalent spatial topologies given in 
Table \ref{Tb-2-Orient_and_Non_orient}, namely the orientable slab space ($E_{16}$)
and the non-orientable slab space with flip ($E_{17})$.

\subsection{The point charge case}
\label{Subsec-charge}

We first consider a  nonrelativistic test particle with
charge $q$ and mass $m$ locally subjected to
vacuum fluctuations of the electric field ${\bf E}({\bf x}, t)$
in the topologically nontrivial spacetime
manifold
equipped with the spatially flat Friedmann-Robertson Walker (FRW) metric
\begin{equation}
\label{metric}
ds^2 = dt^2 -a^2(t)(dx^2 + dy^2 + dz^2)\, ,
\end{equation}
which is the particular case of equation (\ref{RWmetric}) with $k=0$ and Cartesian
instead of spherical coordinates. The covariant equation of motion
is~\cite{Weinberg:1972kfs}
\begin{equation}\label{eqmotioncov}
\frac{Du^{\mu}}{d\tau} \overset{\text{\scriptsize def}}{=} \frac{du^{\mu}}{d\tau}
+ \Gamma^{\mu}_{\alpha \beta}u^{\alpha} u^{\beta}  = \frac{f^{\mu}}{m},
\end{equation}
where $u^{\mu}=dx^{\mu}/d\tau$ is the particle's four-velocity, $m$ is its mass,
$\tau$ is its proper time and $f^{\mu}$ is the nongravitational four-force acting on it.
Since we are interested in the motion of a charged particle in an electromagnetic field,
the four-force is   $f^{\mu} = q F^{\mu \nu}u_{\nu}$ where $F^{\mu\nu}$ is the
electromagnetic field tensor.

In the nonrelativistic case, in which the particle's proper time is indistinguishable
from the cosmic time $t$, the equation of motion for the point charge
becomes~\cite{Bessa:2008pr}
\begin{equation}\label{eqmotion}
\frac{d{\bf u}}{dt} + 2\frac{\dot a}{a}{\bf u} = \frac{q}{m} \,{\bf E}({\bf x}, t)\,,
\end{equation}
which can be written as
\begin{equation}\label{eqmotion-simpl}
\frac{1}{a^2}\frac{d}{dt}( a^2{\bf u}) =\frac{q}{m} \,{\bf E}({\bf x}, t)\, ,
\end{equation}
where ${\bf E} =(E^1,E^2,E^3)$ with $E^i=F^{i0}$.
After integration this yields
\begin{equation}\label{eqmotion-integr}
 a^2(t){\bf u}({\bf x},t) = \frac{q}{m}
 \int_{t_i}^t  a^2(t^{\prime}){\bf E}({\bf x}, t^{\prime})dt^{\prime}\,.
\end{equation}
where we have assumed that the particle is initially at rest: ${\bf u}({\bf x},t_i)=0$.
Since proper (physical) distances $d$ at time $t$ are related to coordinate  distances
$r$ by $d=a(t) r$, the proper (physical) velocity $\bf v$ is related to the coordinate
velocity $\bf u$ at time $t$ by
\begin{equation}\label{physveloc}
 {\bf v}({\bf x},t)  = a(t){\bf u}({\bf x},t)\,.
\end{equation}
Therefore, in terms of the physical velocity, Eq.~(\ref{eqmotion-integr}) becomes
\begin{equation}\label{eqmotion-integr-phys-veloc}
 {\bf v}({\bf x},t)  = \frac{q}{m}\frac{1}{a(t)}
 \int_{t_i}^t  a^2(t^{\prime}){\bf E}({\bf x}, t^{\prime})dt^{\prime}\,,
\end{equation}
from which the one can write the dispersion of each velocity component as%
\footnote{By definition, $\,\langle \Delta v^i(\mathbf{x}, t)^2 \rangle =
\langle  v^i(\mathbf{x}, t)^2  \rangle -\langle  v^i(\mathbf{x}, t)  \rangle^2$.}
\begin{equation}\label{phys-veloc-dispersion}
\Bigl\langle \Delta v^i({\bf x},t)^2 \Bigr\rangle =\frac{q^2}{m^2 a^2(t)}
\int_{t_i}^t\int_{t_i}^t a^2(t^{\prime})\,a^2(t^{\prime\prime}) \, \Bigl\langle
E^i({\bf x}, t^{\prime})E^i({\bf x},
 t^{\prime\prime}) \Bigr\rangle_{FRW} \, dt^{\prime}dt^{\prime\prime}\,,
\end{equation}
where $i=1,2,3$ for the corresponding directions $x,y,z$.
\footnote{For any three-vector $\bf b$ we write either ${\bf b}=(b^1,b^2,b^3)$
or ${\bf b} = (b_x,b_y,b_z)$ according to convenience.}
As in Ref.~\cite{Yu:2004tu,Lemos:2020ogj},
here we assume that $\bf x$ is constant, meaning that in the time scales of interest
the particle's position essentially does not change.

\subsection{Velocity dispersion in terms of conformal time}
\label{Conformaltime}

The Friedmann-Robertson-Walker correlation function that appears in Eq.~(\ref{phys-veloc-dispersion})
can be more easily computed in terms of the Minkowski spacetime correlation function by making use
of how the electromanetic field changes under a conformal transformation.
Indeed, in terms of the conformal
time $\eta$ defined by $dt = a(t) d\eta$ the FRW metric becomes 
%
\begin{equation}\label{metric-conformal}
ds^2 = a^2(d\eta^2 - dx^2 -  dy^2 - dz^2)\,.
\end{equation}
In these coordinates the electromagnetic field  tensor in  FRW spacetime is related
to the one in Minkowski spacetime $ M$ by~\cite{Wald:1984rg,Bessa:2008pr}
\begin{equation}\label{electromg-conformal-transf}     
F^{\mu \nu}({\bf x}, \eta )_{F\!RW} = a^{-4} F^{\mu \nu}({\bf x}, \eta )_{M}\,.
\end{equation}
Taking this equation into account, noting further that the coordinate change $t \to \eta$
implies
\begin{equation}\label{F01-transf}
F^{i0}({\bf x}, \eta )_{F\!RW} = a^{-1} F^{i0}({\bf x}, t)_{F\!RW}\, ,
\end{equation}
and changing the integration variable to $\eta$ using $dt = a\, d\eta$,
Eq.~(\ref{phys-veloc-dispersion}) reduces to
\begin{equation}\label{veloc-dispersion-eta}  
\Bigl\langle \Delta v^i({\bf x},t)^2 \Bigr\rangle =\frac{q^2}{m^2 a^2(t)}
\int_{\eta_i}^{\eta}
\int_{\eta_i}^{\eta} \Bigl\langle E^i({\bf x}, \eta^{\prime})E^i({\bf x}, \eta^{\prime\prime})
\Bigr\rangle_M \, d\eta^{\prime}d\eta^{\prime\prime}\,.
\end{equation}
Therefore, in order to compute the velocity dispersion all one needs is the correlation
function in Minkowski spacetime with the time coordinate replaced by $\eta$.
The result can be expressed in terms of the cosmic time $t$ as long as the scale factor
is known as a function of $t$ or $\eta$.

Since the correlation function in equation~\eqref{veloc-dispersion-eta} is in Min-kowski
spacetime $M$, it depends on the topology of the spatial section as discussed in
\cite{Bessa:2019aar,Lemos:2020ogj}. Then the above result for
the velocity dispersion~~\eqref{veloc-dispersion-eta} in FRW spacetime depends on
the topology of the spatial sections $M_3$ of the FRW spacetime, whose set of
possible nonequivalent topologies is identical to the corresponding set for
the spatial section $M_3$ of Minkowski spacetime $M$.
In the next section we shall use this result to derive the velocity dispersion,
 and thus explicit expressions for a velocity dispersion orientability indicator
for manifolds endowed with non-orientable topology $E_{17}$, and
for its orientable counterpart  $E_{16}$.

\section{Orientability indicators for  $\mathbf{E_{17}}$ and  $\mathbf{E_{16}}$
 spatial topologies} \label{disp-particleE17}

In this section, we compute the orientability indicators
for a FRW spacetime with flat  spatial sections equipped with the non-orientable
$E_{17}$ topology.
The corresponding results for spatial sections endowed
with $E_{16}$ topology follow from those for  $E_{17}$ with no
need of additional calculations.

Following Yu and Ford~\cite{Yu:2004tu}, we assume that the electric field
$\mathbf{E}$ is a sum of classical  $\mathbf{E}_c$ and quantum  $\mathbf{E}_q$
parts. Since there are no quantum fluctuations of  ${\bf E}_c$  and
$\langle {\bf E}_q\rangle =0$, the two-point function
$\bigl\langle E^i({\bf x}, \eta^{\prime})E^i({\bf x}, \eta^{\prime\prime})
\bigr\rangle_M$ in equation~\eqref{veloc-dispersion-eta} involves only
the quantum part of the electric field~\cite{Yu:2004tu}.

It can be shown \cite{Birrell:1982ix} that locally
\begin{equation}\label{eqdif-00}
\Bigl \langle E^i({\bf x}, \eta)E^i({\bf x}', \eta')\Bigl \rangle =
\frac{\partial }{\partial x_i}
\frac{\partial}
{\partial {x'}_i}D({\bf x}, \eta ; {\bf x}', \eta') - \frac{\partial }{\partial \eta }
\frac{\partial}
{\partial \eta'}D ({\bf x }, \eta; {\bf x'}, \eta')
\end{equation} 
The topology of the spatial section $M _3$ is (globally) taken into account as follows.
When  $M_3$ is simply-connected
the Hadamard function $D({\bf x}, \eta;{\bf x}', \eta')$ is given by
\begin{equation}\label{eqren}
D_0({\bf x}, \eta; {\bf x}', \eta') = \frac{1}{4\pi^2(\Delta \eta^2 -
|\Delta \mathbf{x}|^2)} \,.
\end{equation}
The subscript $0$ indicates standard Minkowski spacetime $M$,
$\Delta \eta = \eta - \eta'$ and  $|\Delta \mathbf{x}| \equiv r $
is the spatial separation for topologically trivial Minkowski
spacetime:
\begin{equation}\label{separation-trivial}
r^2 = (x-x')^2 + (y-y')^2 + (z - z')^2  \,.
\end{equation}
However, when Minkowski spacetime in endowed with a topologically nontrivial
spatial section, the {spatial separation} $r^2$ takes a different form that
captures the periodic boundary conditions imposed on the covering space
$\mathbb{E}^{3}$ by the covering group  $\Gamma$, which characterize the
spatial topology. In Table~\ref{Tb-Spatial-separation}
we collect the {spatial separations} for the topologically non-homeomorphic
Euclidean spaces we shall address in this paper.%
\footnote{The reader is referred to Refs.~\cite{LACHIEZEREY1995135,%
Riazuelo:2003ud,Fujii:2011ga} for pictures of the fundamental cells and
further properties of all possible three-dimensional Euclidean topologies.}

\begin{table}[h!]
\begin{tabular}{*2{l|c}}
\hline\hline
Spatial topology       & $\quad$ Spatial separation $r^2$ for Hadamard function  \\
\hline
$E_{16}$ - Slab space           & $(x - x'- n_x\, L)^2 + (y - y')^2 + (z - z')^2 $ \\ 
\hline
$E_{17}$ - Slab space with flip &
\hspace{-.25cm}$\left(x-x'-n_x \,L \right)^{2}+\left(y -(-1)^{n_x} y'\right)^{2}
+ \left(z - z' \right)^{2}$   
\\  \hline
\end{tabular}
\caption{Spatial separation in Hadamard function for the  multiply-connected flat
orientable ($E_{16}$) and its non-orientable counterpart ($E_{17}$) quotient Euclidean
manifolds. The topological compact length is denoted by $L$.
The numbers $n_x$ are integers  and run from $-\infty$ to $\infty$.
For each multiply-connected topology, when  $n_x=0$ we recover the
spatial separation for the simply-connected Euclidean $3-$space.} %
\label{Tb-Spatial-separation}
\end{table}

To obtain the correlation function for the electric field
that is required to compute the velocity dispersion~\eqref{veloc-dispersion-eta} for
slab space with flip $E_{17}$,  we replace in Eq.~\eqref{eqdif-00} the Hadamard
function $D({\bf x}, \eta; {\bf x}', \eta')$ by its renormalized version
given by~\cite{Bessa:2019aar}
\begin{equation}\label{Hadamard-ren}   
D_{ren}({\bf x}, \eta; {\bf x}', \eta') =
D({\bf x}, \eta; {\bf x}', \eta') - D_0({\bf x}, \eta; {\bf x}', \eta')
=\sum\limits_{{n_x=-\infty}}^{{\infty\;\;\prime}}\frac{1}{4\pi^2(\Delta \eta^2 - r^2)}\,,
\end{equation}
where  here and in what follows $\sum_{}^{\;'}$ indicates that
the Minkowski contribution term $n_x = 0$ is excluded from the summation,
$\Delta \eta = \eta -\eta^{\prime}$, and the spatial separation $r$ for $E_{17}$
is given in Table~\ref{Tb-Spatial-separation}.
The term  with $n_x=0$ that would be present in  the sum~\eqref{Hadamard-ren} is the Hadamard function
$D_{0}({\bf x}, \eta; {\bf x}', \eta')$ for  Minkowski spacetime with simply-connected
spatial section $\mathbb{E}^3$.
This term  has been subtracted out from the sum
because it gives rise to an infinite contribution
to the velocity dispersion~\cite{Bessa:2019aar,Lemos:2020ogj}.

Thus, from equation (\ref{eqdif-00}) the renormalized correlation
functions
\begin{equation}\label{correlation-i-E17}
\Bigl \langle E^i({\bf x}, \eta)E^i({\bf x}', \eta')\Bigr \rangle_{ren}
=\frac{\partial }{\partial x_i} \frac{\partial}
{\partial {x'}_i}D_{ren}({\bf x}, \eta; {\bf x}', \eta') - \frac{\partial }{\partial \eta}
\frac{\partial}
{\partial \eta'}D_{ren} ({\bf x }, \eta; {\bf x'}, \eta')\,,
\end{equation}
where $D_{ren}({\bf x}, \eta; {\bf x}', \eta')$ depends on the
spatial topology through $r$   according to (\ref{Hadamard-ren}) and
Table~\ref{Tb-Spatial-separation}.

{}From equations (\ref{correlation-i-E17}), (\ref{Hadamard-ren}) with $r$ given
in Table~\ref{Tb-Spatial-separation} the electric field correlation functions
for $E_{17}$ topology are found to be given by
\begin{eqnarray}\label{correlation-x-E17}   
&&\Bigl\langle E_x({\bf x}, \eta )E_x({\bf x}', \eta')\Bigl\rangle_{ren}^{E_{17}}  =
\sum\limits_{{n_x=-\infty}}^{{\infty\;\;\prime}}
\frac{\Delta \eta^2 + r^2 -2r_x^2}{\pi^2 [\Delta \eta^2 - r^2]^3},
\\ \label{correlation-y-E17}
&&\Bigl\langle E_y({\bf x}, \eta )E_y({\bf x}', \eta')\Bigl\rangle_{ren}^{E_{17}}  =
\sum\limits_{{n_x=-\infty}}^{{\infty\;\;\prime}}\bigg[
\frac{\left( 3-(-1)^{n_x}\right) \Delta \eta^2}{2\pi^2 [\Delta \eta^2 - r^2]^3}
+\frac{\left( 1+ (-1)^{n_x}\right) r^2
- 4(-1)^{n_x} r_y^2} {2\pi^2 [\Delta \eta^2 - r^2]^3}\bigg],
\\ \label{correlation-z-E17}
&&\Bigl\langle E_z({\bf x}, \eta )E_z({\bf x}', \eta' )\Bigl\rangle_{ren}^{E_{17}}  =
\sum\limits_{{n_x=-\infty}}^{{\infty\;\;\prime}}
\frac{\Delta \eta^2 + r^2 -2r_z^2}{\pi^2 [\Delta \eta^2 - r^2]^3},
\end{eqnarray}
where $\Delta \eta = \eta -\eta^{\prime}$ and
\begin{equation}\label{r-components-E17}
r_x = x-x^{\prime}-n_xL,\;\;\;\; r_y = y-(-1)^{n_x}\, y^{\prime},\;\;\;\;
 r_z = z-z^{\prime}, \;\;\;\;r= \sqrt{r_x^2 + r_y^2+ r_z^2}\,.
\end{equation}

The orientability indicator $\mbox{\large $I$}_{v^2_i}^{E_{17}}$ that we will
consider here is defined by replacing
the electric field correlation functions in Eq.~\eqref{veloc-dispersion-eta} by
their renormalized counterparts~(\ref{correlation-i-E17}) in which $r$ is given
in Table~\ref{Tb-Spatial-separation}, namely
\begin{equation}   \label{indicator-E17}
 \mbox{\large $I$}_{v^2_i}^{E_{17}} ({\bf x},t) = \frac{q^2}{m^2a^2(t)}
 \int_0^{\eta}\int_0^{\eta}\Bigl\langle E^i({\bf x}, \eta') E^i({\bf x}, \eta'')\Bigr \rangle_{ren}^{E_{17}}\, d\eta' d\eta''\,.
\end{equation}
Equation~(\ref{Hadamard-ren}) makes it is clear  that the orientability indicator
$\mbox{\large $I$}_{v^2_i}^{E_{17}}$ is the difference between the velocity
dispersion in $E_{17}$ and the one in Minkowski with trivial (simply connected) topology

\subsection{Orientability indicator: general definition}
\label{Indicatorgeneral}

For later use and for the sake of generality, some words of clarification are
in order at this point before proceeding to the
calculation of the components of the statistical indicator.
From equations~~\eqref{veloc-dispersion-eta} and~\eqref{Hadamard-ren}
a definition of the orientability indicator for a general multiply-connected
flat topology can be written in
the form \cite{Lemos:2020ogj}
\begin{equation}  \label{new-ind}     
 \mbox{\large $I$}_{v^2_i}^{MC}
= \Bigl\langle\Delta v_i^2 \Bigr \rangle^{MC}
- \;\,\Bigl \langle\Delta v_i^2 \Bigr \rangle^{SC} ,
\end{equation}
where  $\bigl\langle\Delta v_i^2 \bigr \rangle$  is the
mean square velocity dispersion, and the superscripts $MC$ and $SC$
stand for multiply- and simply-connected topologies, respectively.
The right-hand side of~(\ref{new-ind}) is defined  in the following
way: one first takes the difference
of the two terms with ${\bf x}^{\prime} \neq  {\bf x}$ and then  sets
${\bf x}^{\prime} =  {\bf x}$. Since $\mbox{\large $I$}_{v^2_i}^{MC}$ is not
simply the velocity dispersion $\bigl\langle\Delta v_i^2 \bigr\rangle^{MC}$ but
the difference~\eqref{new-ind}, the possibility
that it takes negative values
should not be a matter of worry,
a point that does not seem to have been  appreciated in some previous works in which
this indicator was implicity used~\cite{Yu:2004tu,Bessa:2019aar,Yu:2004gs,%
Ford:2005rs,Yu:2006tn,Seriu_2008,Parkinson:2011yx,DeLorenci:2016jhd,jun2004vacuum}
together with the particular assumption that 
the second term vanishes.\footnote{It is  experimentally and theoretically unsettled whether 
the simply-connected term on the right-hand side of (\ref{new-ind}) vanishes
or not.
We adopt the general view that it is nonzero \cite{Lemos:2020ogj}. 
It is only under the very special assumption that it vanishes that one
stumbles upon the counterintuitive negative values for mean square velocities often found in the literature
\cite{Yu:2004tu,Bessa:2019aar,jun2004vacuum,Ford:2005rs,Seriu_2008,Parkinson:2011yx,DeLorenci:2016jhd,Yu:2004gs}.}

At first sight, the indicator \eqref{new-ind} does not seem measurable because it involves the difference of quantities associated with
two different spacetimes, but spacetime is unique. However, $\mbox{\large $I$}_{v^2_i}^{MC}$ is accessible by measurements performed only in our  spacetime, which is to be tested for multiple connectedness, by the following procedure. First one would measure the  velocity correlation function
$\bigl\langle \Delta v_i({\bf x},t) \Delta v_i({\bf x}^{\prime},t)\bigr\rangle^{exp}$ for
${\bf x} \neq {\bf x}^{\prime}$, then  one would subtract out the correlation
function $\bigl\langle \Delta v_i({\bf x},t) \Delta v_i({\bf x}^{\prime},t)\bigr\rangle^{SC}$ that can be {\it  theoretically computed}  for
${\bf x} \neq {\bf x}^{\prime}$
for the corresponding  topologically trivial (simply-connected) FRW spacetime, just as was done in the Appendix of Ref. \cite{Lemos:2020ogj} in the case of the Minkowski spacetime.\footnote{We do not present here the results for the topologically trivial FRW spacetime for the sake of brevity.}
Finally, the corresponding curve for the difference \eqref{new-ind}     as a function of time would be plotted in the coincidence limit
${\bf x} = {\bf x}^{\prime}$ (see, for example, the figures in Section \ref{casestudy}). This approach is similar to one used in  cosmic crystallography, which is intended to detect cosmic
topology from the distribution of  discrete cosmic sources~\cite{Lehoucq:1996qe}.
A topological signature of any multiply connected
$3-$manifold of constant curvature is given by a constant times the
difference $\Phi_{exp}^{MC}(s_i) - \Phi^{SC}_{exp}(s_i)$ of the
expected pair separation histogram (EPSH) corresponding to the multiply
connected manifold and the EPSH for the underlying simply connected
covering manifold~\cite{Gomero:2001gq,Gomero:1998dz}, whose expression
{\it can be derived in analytical form}~\cite{Gomero:2001gq,Reboucas:2000du}.

It should be noticed that, although with  the present techniques the second term on 
the right-hand side of \eqref{new-ind} cannot be directly computed at 
${\bf x}={\bf x}^{\prime}$, we can nonetheless presume that a rigorous theoretical 
treatment may allow a consistent determination of this term. From our current knowledge 
of the theory, one can also reasonably expect that a correct renormalization procedure 
will not significantly change the qualitative features of the two terms, at least not 
to the extent that it would affect comparison with observational data.

\subsection{Orientability indicators for charged point particle}
\label{Indicatorpointcharge}

Let us return to the  calculation of the 
components of the orientability indicator  for ${E_{17}}$
topology.
Since the correlation functions~(\ref{correlation-x-E17}) to~(\ref{correlation-z-E17})
depend on $\eta$ and $\eta'$ only through their difference, the changes of integration
variables $\eta_1 = \eta^{\prime} - \eta_i$ and $\eta_2 = \eta^{\prime\prime} - \eta_i$
in Eq.~(\ref{veloc-dispersion-eta}) allow the components of the velocity dispersion
to be computed with the help of the integrals~\cite{Bessa:2019aar}
\begin{eqnarray}\label{integral1}  
{\cal I} & = &\int_0^{\Delta \eta} \! \int_0^{\Delta \eta} d\eta_1d\eta_2
\frac{1}{[ (\eta_2 - \eta_1)^2 -r^2]^3}  \nonumber \\
& = &\frac{\Delta \eta}{16r^5 (\Delta \eta^2-r^2)}\bigg\{4r\Delta \eta - 3(r^2-\Delta \eta^2)
\ln \frac{(r-\Delta \eta)^2}{(r+\Delta \eta)^2} \bigg\}
\end{eqnarray}
and
\begin{eqnarray}\label{integral2}
{\cal J} & = & \int_0^{\Delta \eta} \int_0^{\Delta \eta} d\eta_1 d\eta_2  \frac{(\eta_2
- \eta_1)^2}{[(\eta_2 - \eta_1)^2 -r^2]^3}
\nonumber \\
& = &\frac{\Delta \eta}{16r^3(\Delta \eta^2-r^2)}\bigg\{ 4r\Delta \eta + (r^2-\Delta \eta^2)
\ln  \frac{(r-\Delta \eta )^2}{(r+\Delta \eta )^2} \bigg\}\,,
\end{eqnarray}
in which $\Delta \eta = \eta - \eta_i$.

Inserting  equations~(\ref{correlation-x-E17}) to (\ref{integral2}) into
Eq.~(\ref{veloc-dispersion-eta})
and taking the coincidence limit  ${\bf x}^{\prime} \to {\bf x}$ we find
\begin{eqnarray}\label{dispersion-x-E17}   
 {\mbox{\large $I$}_{v^2_x}^{E_{17}}}({\bf x},t)_{_{\mbox{\scriptsize $FRW$}}} & = &
\sum\limits_{{n_x=-\infty}}^{{\infty\;\;\prime}}
\frac{q^2 \Delta\eta}{16\pi^2 m^2 r^5 (\Delta\eta^2 - r^2) a^2(t)}
 \bigg\{ 4r\Delta\eta ({\bar r}_x^2 + r^2) \nonumber \\
&  & \hspace*{2.5cm} +(\Delta\eta^2-r^2)(3{\bar r}_x^2 -r^2)\ln \frac{(r-\Delta\eta )^2}{(r+\Delta\eta )^2} \bigg\},
 \\
\label{dispersion-y-E17}
{\mbox{\large $I$}_{v^2_y}^{E_{17}}}({\bf x},t)_{_{\mbox{\scriptsize $FRW$}}} & = &
 \sum\limits_{{n_x=-\infty}}^{{\infty\;\;\prime}} \frac{q^2 \Delta\eta}{32\pi^2 m^2 r^5
 (\Delta\eta^2 - r^2) a^2(t)}
\bigg\{ 4r\Delta\eta ({\bar r}_y^2 + (3-(-1)^{n_x})r^2) \nonumber \\
 & & \hspace*{2.5cm} + (\Delta\eta^2-r^2)[3{\bar r}_y^2 -(3-(-1)^{n_x})r^2]
\ln \frac{(r-\Delta\eta )^2}{(r+\Delta\eta )^2} \bigg\}\,, \\
\label{dispersion-z-E17}
{\mbox{\large $I$}_{v^2_z}^{E_{17}}}({\bf x},t)_{_{\mbox{\scriptsize $FRW$}}} & = &
\sum\limits_{{n_x=-\infty}}^{{\infty\;\;\prime}}\frac{q^2 \Delta\eta}{16\pi^2 m^2 r^5
(\Delta\eta^2 - r^2)a^2(t)}
\bigg\{ 4r\Delta\eta ({\bar r}_z^2  + r^2) \nonumber \\
& & \hspace*{2.5cm} + (\Delta\eta^2-r^2)(3{\bar r}_z^2 -r^2)
\ln \frac{(r-\Delta\eta )^2}{(r+\Delta\eta )^2} \bigg\}\,, 
\end{eqnarray}
where
\begin{eqnarray}\label{r-ri-coincidence-r}
r & = & \sqrt{n_x^2L^2+2(1-(-1)^{n_x})y^2}, \\
\label{r-ri-coincidence-r-bar-x}
{\bar r}_x^2 & = & r^2- 2r_x^2 = -n_x^2 L^2 + 2(1-(-1)^{n_x})y^2, \\
\label{r-ri-coincidence-r-bar-y}
{\bar r}_y^2 & = & (1+ (-1)^{n_x}) r^2 - 8 (-1)^{n_x} (1-(-1)^{n_x})y^2 \nonumber \\
&  = &  (1+ (-1)^{n_x})n_x^2 L^2 + 8 (1-(-1)^{n_x})y^2,\\
\label{r-ri-coincidence-r-bar-z}
{\bar r}_z^2 & = & r^2- 2r_z^2 = r^2\,,
\end{eqnarray}
with the use of equations~(\ref{r-components-E17}) in the coincidence limit.

Note that the velocity dispersion depends not only on the time interval
$\Delta t =t -t_i$ (or $\Delta \eta = \eta - \eta_i$) but also on $t$ (or $\eta$)
itself. This was to be expected because in a dynamical universe  invariance under
time translations is lost.

The orientability indicator for $E_{16}$ follows easily from the results for $E_{17}$.
The factors of $(-1)^{n_x}$ that appear in equations \eqref{dispersion-x-E17}
to~\eqref{dispersion-z-E17}
arise from derivatives with respect to $y^{\prime}$ in the separation $r$ given in
Table~\ref{Tb-Spatial-separation}. 
Hence, the results for $E_{16}$ are immediately
obtained from those for $E_{17}$ by simply replacing $(-1)^{n_x}$ by $1$ everywhere
in Eqs.~(\ref{dispersion-x-E17}) to (\ref{dispersion-z-E17}). This leads to
\begin{eqnarray}\label{dispersion-orientable-x-E16}   
 {\mbox{\large $I$}_{v^2_x}^{E_{16}}}({\bf x},t)_{_{\mbox{\scriptsize $FRW$}}} & = &
-\frac{q^2 \Delta \eta}{4\pi^2 m^2 a^2(t)} \sum\limits_{{n_x=-\infty}}^{{\infty\;\;\prime}}
\frac{1}{n^3L^3} \ln \frac{(n_x L-\Delta \eta )^2}{(n_x L+\Delta \eta )^2}\,, \\
\label{dispersion-orientable-yz-E16}
 {\mbox{\large $I$}_{v^2_y}^{E_{16}}}({\bf x},t)_{_{\mbox{\scriptsize $FRW$}}} & = &
 {\mbox{\large $I$}_{v^2_z}^{E_{16}}}({\bf x},t)_{_{\mbox{\scriptsize $FRW$}}} \nonumber \\
& = & \frac{q^2 \Delta \eta}{8\pi^2 m^2 a^2(t)}
\sum\limits_{{n_x=-\infty}}^{{\infty\;\;\prime}}
\bigg\{ \frac{4\Delta \eta}{n_x^2L^2(\Delta \eta^2- n_x^2 L^2)} + \frac{1}{n_x^3L^3}
\ln \frac{(n_xL-\Delta \eta )^2}{(n_xL + \Delta \eta )^2} \bigg\}\,.
\end{eqnarray}

\section{NON-ORIENTABILITY WITH POINT ELECTRIC DIPOLE}
\label{Dipole-motion}

A noteworthy  outcome of the previous section is that %
the time evolution of the orientability indicator for a charged particle
can be  used to locally differentiate  an orientable ($E_{16}$) from a
non-orientable ($E_{17}$) spatial section of Minkowski spacetime.
However,  it cannot be used to decide whether a given $3-$space manifold per se 
is or not orientable. As shown in \cite{Lemos:2020ogj}, the spatial orientability
of Minkowski spacetime in itself can be ascertained in principle by the motions of a
point electric dipole.
Therefore, it is reasonable to expect
that the orientability indicator for a dipole can potentially bring about unequivocal information
regarding non-orientability of the spatial sections of the spatially flat FRW spacetime.
To examine this issue  we now turn our attention 
to topologically induced motions of an electric dipole under quantum vacuum electromagnetic
fluctuations.

The spatial components of the four-force on a point electric dipole  are
$f^i = p^j\partial_j E^i$ where ${\bf p}=(p^1,p^2,p^3)$ is the electric dipole
moment vector. Since the dipole is taken to be a bound system, it is not affected
by the expansion of the universe, which means that the dipole moment $\bf p$
is a constant vector.
Under the same assumptions as made for the point particle, replacing $qE^i$ by
$p^j\partial_j E^i$ and following the same steps that led from (\ref{eqmotion})
to (\ref{phys-veloc-dispersion}),  the  mean squared velocity in each of the three
independent directions $i = x, y, z$ is given by
\begin{equation}\label{eqdispersion-dipole}
\Bigl \langle\Delta v^i({\bf x},t)^2 \Bigr \rangle = \frac{p^jp^k}{m^2a^2(t)}
\int_{t_i}^t\int_{t_i}^t a^2(t^{\prime})a^2(t^{\prime\prime})
\Bigl\langle \bigl(\partial_j E^i({\bf x}, t')\bigr)
\bigl(\partial_k E^i({\bf x}, t'')\bigr)\Bigr\rangle_{FRW}\, dt' dt''\,,
\end{equation}
which can be conveniently rewritten as
\begin{equation}\label{eqdispersion-dipole-rewritten}
\Bigl \langle\Delta  v^i({\bf x},t)^2 \Bigl\rangle = \lim_{{\bf x}^{\prime}
\to {\bf x}}\frac{p^jp^k}{m^2a^2(t)}\int_{t_i}^t\int_{t_i}^t a^2(t^{\prime})
a^2(t^{\prime\prime})
\partial_j\partial_k^{\,\prime} \Bigl\langle  E^i({\bf x}, t')
 E^i({\bf x}', t'')\Bigr \rangle_{FRW} dt' dt'' \,,
\end{equation}
where $\partial_l^{\,\prime}= \partial/\partial x_l^{\prime}$ and the
summation convention applies only to repeated upper and lower indices.

Now we proceed to the computation of  the orientability indicator for a point dipole
in spaces $E_{17}$ and $E_{16}$. The space $E_{17}$ has two topologically
special directions:    
the compact $x$-direction  and the flip $y$-direction associated 
with the non-orientability of $E_{17}$.
To probe the non-orientability of $E{_{17}}$ by means
of stochastic motions it seems most promising to choose  a dipole oriented
in the $y$-direction, since the orientation of the dipole would also be flipped
upon every displacement by the topological length  $L$ along the compact
direction~\cite{Lemos:2020ogj}.

For a dipole oriented along the $y$-axis we have ${\bf p}=(0,p,0)$
and
\begin{equation}\label{eqdispersion-dipole-y}
\Bigl \langle\Delta v_x({\bf x},t)^2 \Bigr\rangle^{(y)} = \lim_{{\bf x}^{\prime}
\to {\bf x}}\frac{p^2}{m^2a^2(t)}\int_{t_i}^t\int_{t_i}^t a^2(t^{\prime})
a^2(t^{\prime\prime})
\partial_y\partial_{y^{\prime}} \Bigl\langle  E_x({\bf x}, t')
 E_x({\bf x}', t'')\Bigr\rangle_{FRW}\, dt' dt'',
\end{equation}
where the superscript within parentheses  indicates the dipole's orientation.
We now proceed as in the case of the charged particle and rewrite the above
integral in terms of the conformal time and the correlation function in
Minkowski spacetime. It follows that the above equation takes the form
\begin{equation}\label{eqdispersion-dipole-y-conformal}
\Bigl \langle\Delta v_x({\bf x},t)^2\Bigr\rangle^{(y)} =\lim_{{\bf x}^{\prime}
\to {\bf x}}\frac{p^2}{m^2a^2(t)}
\times \int_{\eta_i}^{\eta}\int_{\eta_i}^{\eta} \partial_y\partial_{y^{\prime}}
\Bigl\langle  E_x({\bf x}, \eta')
 E_x({\bf x}', \eta'')\Bigr\rangle_M\, d\eta' d\eta''\,.
\end{equation}

Upon replacing the eletric-field correlation function by its renormalized version
given by Eq.~(\ref{correlation-x-E17}), the above equation yields the orientability
indicator in the $x$-direction for $E_{17}$:
\begin{equation}\label{eqdispersion-dipole-yx}
{\mbox{\large $I$}_{v^2_x}^{E_{17}}}({\bf x},t)_{_{\mbox{\scriptsize $FRW$}}}^{^{\mbox{\scriptsize $(y)$}}}
 =  \lim_{{\bf x}^{\prime}
\to {\bf x}}\frac{p^2}{\pi^2m^2a^2(t)} \sum\limits_{{n_x=-\infty}}^{{\infty\;\;\prime}}
\int_{\eta_i}^{\eta}\int_{\eta_i}^{\eta}
\partial_y\partial_{y^{\prime}}  \frac{\Delta \eta^2 + r^2 -2r_x^2}{({\Delta \eta}^2 - r^2)^3}\,.
\end{equation}
where  $\Delta \eta = \eta'-\eta''$, while  $r_x$ and $r$ are defined by Eq.~(\ref{r-components-E17}).
Making use of
\begin{equation}\label{partial-y-yprime-yx}
\partial_y\partial_{y^{\prime}} \frac{\Delta \eta^2 + r^2 -2r_x^2}{({\Delta \eta}^2 - r^2)^3}
= -4(-1)^{n_x} \bigg[ \frac{2}{({\Delta \eta}^2 -r^2)^3}
+ 3 \frac{r^2 - r_x^2+ 6r_y^2}{({\Delta \eta}^2 -r^2)^4}
+ 24 \frac{(r^2 - r_x^2) r_y^2}{({\Delta \eta}^2 -r^2)^5} \bigg].
\end{equation}
we find
\begin{equation}\label{eqdispersion-dipole-yx-final-E17}
{\mbox{\large $I$}_{v^2_x}^{E_{17}}}({\bf x},t)_{_{\mbox{\scriptsize $FRW$}}}^{^{\mbox{\scriptsize $(y)$}}} = -\frac{4p^2}{\pi^2m^2a^2(t)} \sum\limits_{{n_x=-\infty}}^{{\infty\;\;\prime}}(-1)^{n_x} \bigg\{ 2I_1
+ 3 (r^2 - r_x^2 + 6r_y^2)I_2+ 24 (r^2 - r_x^2) r_y^2I_3   \bigg\},
\end{equation}
where~\cite{Lemos:2020ogj} 
\begin{eqnarray}\label{integral-1}
&&I_1  =  {\cal I}   =  \int_0^{\Delta\eta}  \int_0^{\Delta\eta}
\frac{d\eta_1d\eta_2}{[(\eta_2-\eta_1)^2 -r^2]^3}
=    \frac{\Delta\eta}{16} \bigg[ \frac{4\Delta\eta}{r^4 (\Delta\eta^2-r^2)}
+ \frac{3}{r^5} \ln \frac{(r-\Delta\eta )^2}{(r+\Delta\eta )^2} \bigg],\\
&&I_2  =  \int_0^{\Delta\eta}  \int_0^{\Delta\eta} \frac{d\eta_1d\eta_2}{[(\eta_2 -\eta_1)^2 -r^2]^4}
    =  \frac{\Delta\eta}{96}\bigg[ \frac{4\Delta\eta (9r^2-7\Delta\eta^2)}{r^6 (\Delta\eta^2-r^2)^2}
 - \frac{15}{r^7} \ln \frac{(r-\Delta\eta )^2}{(r+\Delta\eta )^2} \bigg],\label{integral-2}\\
&&I_3  =  \int_0^{\Delta\eta}  \int_0^{\Delta\eta} \frac{d\eta_1d\eta_2}{[(\eta_2 -\eta_1)^2 -r^2]^5}
 \nonumber\\
&& \hspace*{3cm} = \frac{\Delta\eta}{768}  \bigg[\frac{105}{r^9} \ln \frac{(r-\Delta\eta )^2}{(r+\Delta\eta )^2}
+ \frac{4\Delta\eta (57\Delta\eta^4- 136r^2\Delta\eta^2 + 87r^4)}
{r^8 (\Delta\eta^2-r^2)^3}  \bigg],\label{integral-3}
\end{eqnarray}
in which $\Delta\eta = \eta - \eta_i$.

Similar  calculations lead to
\begin{eqnarray}\label{eqdispersion-dipole-yy-final-E17}
 {\mbox{\large $I$}_{v^2_y}^{E_{17}}}({\bf x},t)_{_{\mbox{\scriptsize $FRW$}}}^{^{\mbox{\scriptsize $(y)$}}}
& = &  -\frac{2p^2}{\pi^2m^2a^2(t)}\sum\limits_{{n_x=-\infty}}^{{\infty\;\;\prime}} (-1)^{n_x}
\bigg\{ (5-3(-1)^{n_x}) I_1  \nonumber \\
&  & \hspace*{0.2cm} + 6 [r^2 + (7-6(-1)^{n_x})r_y^2] I_2  + 48 [r^2 -(-1)^{n_x}r_y^2]r_y^2I_3 \bigg\}
\end{eqnarray}
and
\begin{equation}
{\mbox{\large $I$}_{v^2_z}^{E_{17}}}({\bf x},t)_{_{\mbox{\scriptsize $FRW$}}}^{^{\mbox{\scriptsize $(y)$}}}   =-\frac{4p^2}{\pi^2m^2a^2(t)} \sum\limits_{{n_x=-\infty}}^{{\infty\;\;\prime}}(-1)^{n_x}\bigg\{ 2I_1
+  +3(r^2 + 6 r_y^2 )I_2 + 24 r^2 r_y^2I_3 \bigg\}.
\label{eqdispersion-dipole-yz-final-E17}
\end{equation}
Since the coincidence limit ${\bf x}^{\prime} \to {\bf x}$ has been taken, it follows
from Eq.~(\ref{r-components-E17}) that in Eqs. (\ref{eqdispersion-dipole-yx-final-E17}) to
(\ref{eqdispersion-dipole-yz-final-E17}) one must put
\begin{equation}
\label{r-rx-ry-coincidence-E17}
 r= \sqrt{n_x^2L^2 + 2(1-(-1)^{n_x})y^2},\; \;
 r_x^2 = n_x^2L^2,\;\;  r_y^2 = 2(1-(-1)^{n_x})y^2.
\end{equation}


For the slab space $E_{16}$ the components of the dipole velocity
dispersion are obtained from those for $E_{17}$  by setting $r_x^2 = r^2, r_y=0 $,
and replacing $(-1)^{n_x}$ by $1$ everywhere. Therefore, we have
\begin{eqnarray}\label{eqdispersion-dipole-yx-final-E16}
{\mbox{\large $I$}_{v^2_x}^{E_{16}}}({\bf x},t)_{_{\mbox{\scriptsize $FRW$}}}^{^{\mbox{\scriptsize $(y)$}}} & = & -\frac{8p^2}{\pi^2m^2a^2(t)}
\sum\limits_{{n_x=-\infty}}^{{\infty\;\;\prime}} I_1 ,\\
\label{eqdispersion-dipole-yy-final-E16}
{\mbox{\large $I$}_{v^2_y}^{E_{16}}}({\bf x},t)_{_{\mbox{\scriptsize $FRW$}}}^{^{\mbox{\scriptsize $(y)$}}} & = & -\frac{4p^2}{\pi^2m^2a^2(t)}
\sum\limits_{{n_x=-\infty}}^{{\infty\;\;\prime}} ( I_1 + 3r^2I_2 ),\\
\label{eqdispersion-dipole-yz-final-E16}
{\mbox{\large $I$}_{v^2_z}^{E_{16}}}({\bf x},t)_{_{\mbox{\scriptsize $FRW$}}}^{^{\mbox{\scriptsize $(y)$}}} & = & -\frac{4p^2}{\pi^2m^2a^2(t)}
\sum\limits_{{n_x=-\infty}}^{{\infty\;\;\prime}} (2I_1 + 3r^2I_2 ),
\end{eqnarray}
in which $r=\vert n_x \vert L$.

It is worth pointing out that the general results obtained for the
orientability indicators are independent of a specific gravitation theory,
they depend only on the assumption that  spacetime is endowed with the
spatially flat metric~\eqref{metric}. We have also found  that the amplitude of the indicator components, both for the charged particle and the electric dipole,   is inversely proportional to $a^2(t)$.
This was not expected from the outset, and had to be derived.

However, the indicators depend on the scale factor,
which is determined by the gravitational theory.
Therefore, in order to get definite expressions for the indicators we shall
consider a specific model within the framework of general relativity.

\section{A case study}\label{casestudy}

The purpose of this section is to study the time evolution of the orientability
indicators for spatially flat expanding  universes with  spatial sections endowed
with either of the  nontrivial topologies  $E_{16}$ or $E_{17}$.
A difficulty arises, however, because the orientability indicators found in the
two previous sections, both for the charged particle and the point dipole, are
expressed in terms of a mixture of the cosmic time $t$ and the conformal
time $\eta$, which makes it difficult to bring out their behavior.
In order to express the orientability indicators exclusively in terms
of $t$ or $\eta$ one has to know the scale factor as a function of
$t$ or $\eta$.

As we have mentioned in the Introduction, the  metric \eqref{metric}
expresses the principle of spatial homogeneity and isotropy along with
the existence of a cosmic time $t$, with the additional observational
input  from the Planck collaboration~\cite{2016,2020}
that provided strong support for a flat $3-$space ($|\Omega_k| < 0.003$).
To study the dynamics of the Universe another assumption  is necessary,
namely that the large scale structure of the Universe is essentially
determined by gravitational interactions, and therefore can be described
by a  metrical theory of gravity, which we  assume to be  general relativity.

These very general assumptions constrain the matter content of the Universe
to be described by a perfect fluid with energy-momentum tensor
\begin{equation}
\label{EM-Tensor}
T_{\mu\nu} = (\rho+p)\,u_\mu u_\nu - p \,g_{\mu \nu}\;,
\end{equation}
where $u_\mu$ is the fluid four-velocity,  $\rho$ is the total energy density and $p$ is the
pressure. In the case of arbitrary space curvature, the Einstein field equations imply the
Friedmann equation
\begin{equation}
\label{Friedmann}
\frac{\dot{a}^2}{a^2} = \frac{8\pi G}{3} \rho -  \frac{k}{a^2},
\end{equation}
where $G$ is the gravitational constant. The conservation law $\nabla_{\mu}T^{\mu\nu}=0$
leads to the fluid equation
\begin{equation}
\label{fluid}
{\dot \rho} + 3\frac{\dot a}{a}(\rho + p) = 0.
\end{equation}

Our  main interest is to show how valuable the orientability indicators are.
Thus, for illustrative purposes  we shall consider a model universe which is
spatially flat and whose  matter content consists of a single-component
barotropic perfect fluid with equation of state $p = w \rho$, where
the pure number $w$ satisfies $\vert w \vert < 1$ .
In this case the expanding solution to the  Friedmann equation~(\ref{Friedmann})
with $k=0$ and the fluid equation (\ref{fluid}) is \cite{Ryden:2002vsj}
\begin{equation}
a=\left(\frac{t}{t_0}\right)^{2/(3+3w)}.
\label{sol_1_componente}
\end{equation}
The age of this universe is
\begin{equation}
t_0=\frac{2}{3(1+w)} H_0^{-1}
\label{idade}
\end{equation}
where $H_0$ is the Hubble constant, the present value of $H= {\dot a}/a$.
According to the latest observational data from the Planck team \cite{2020},
$H_0\simeq 67.37$ km/(s Mpc) or $H_0^{-1}\simeq 14.5$ Gyr and the age of the Universe is
$t_0 \simeq 13.8$ Gyr. In order for equation \eqref{idade} to match these results
we choose the equation of state parameter as $w = -0.299$. Thus, from
Eq.~\eqref{sol_1_componente} we find
\begin{equation}
\label{tempo_conf}
\Delta \eta  =  \int_{t_i}^t \frac{dt^{\prime}}{a(t^{\prime})}
=  \frac{3(1+w)}{1+3w}t_0\left[\left(\frac{t}{t_0}\right)^{\frac{1+3w}{3(1+w)}}
-\left(\frac{t_i}{t_0}\right)^{\frac{1+3w}{3(1+w)}}\right].
\end{equation}
By means of this result the orientability indicators found in
Sections~\ref{disp-particleE17} and \ref{Dipole-motion} are expressed
in terms of the proper time $t$ alone.

It is extremely hard analytically to figure out  how the orientability indicators
behave as functions of cosmic time because their expressions are given by quite
involved infinite sums.  
The visualization of their behavior can be much more easily  accomplished by
numerical plots, which is what we shall  concentrate on in the following. 
For all plots we set $t_i=t_0=1$,  which means that the fluctuations begin to
be measured by the orientability indicators at the same reference instant
at which the scale factor is unity, that is, today.
The compactification length discussed in \eqref{ident-E16}, \eqref{ident-E17} and
Table \ref{Tb-Spatial-separation} is also set equal to unity: $L=1$.
Following \cite{Lemos:2020ogj}, the choice $y=0$, which freezes out the global
inhomogeneity degree of freedom, is made in all plots but
Fig.~\ref{Fig1}(a), in which we take  $y=1/2$ to illustrate the  global
inhomogeneity effect.
The orientability indicators are computed from equation~%
~\eqref{dispersion-x-E17}~--\eqref{dispersion-orientable-yz-E16} for the charged
particle, and equations~\eqref{eqdispersion-dipole-yx-final-E17}~--%
~\eqref{eqdispersion-dipole-yz-final-E16} for the point dipole, as well
as~\eqref{tempo_conf} with $ w =-0.299$.
The infinite sums are rapidly convergent, and the summations are numerically
performed taking $n_x \neq 0$ ranging from $-50$ to $50$.

\subsection{Non-orientability: point charge case}

\begin{figure*}[htpb]
 \begin{center}
\begin{tabular}{c c}
        \resizebox{\halfsize}{!}{\includegraphics{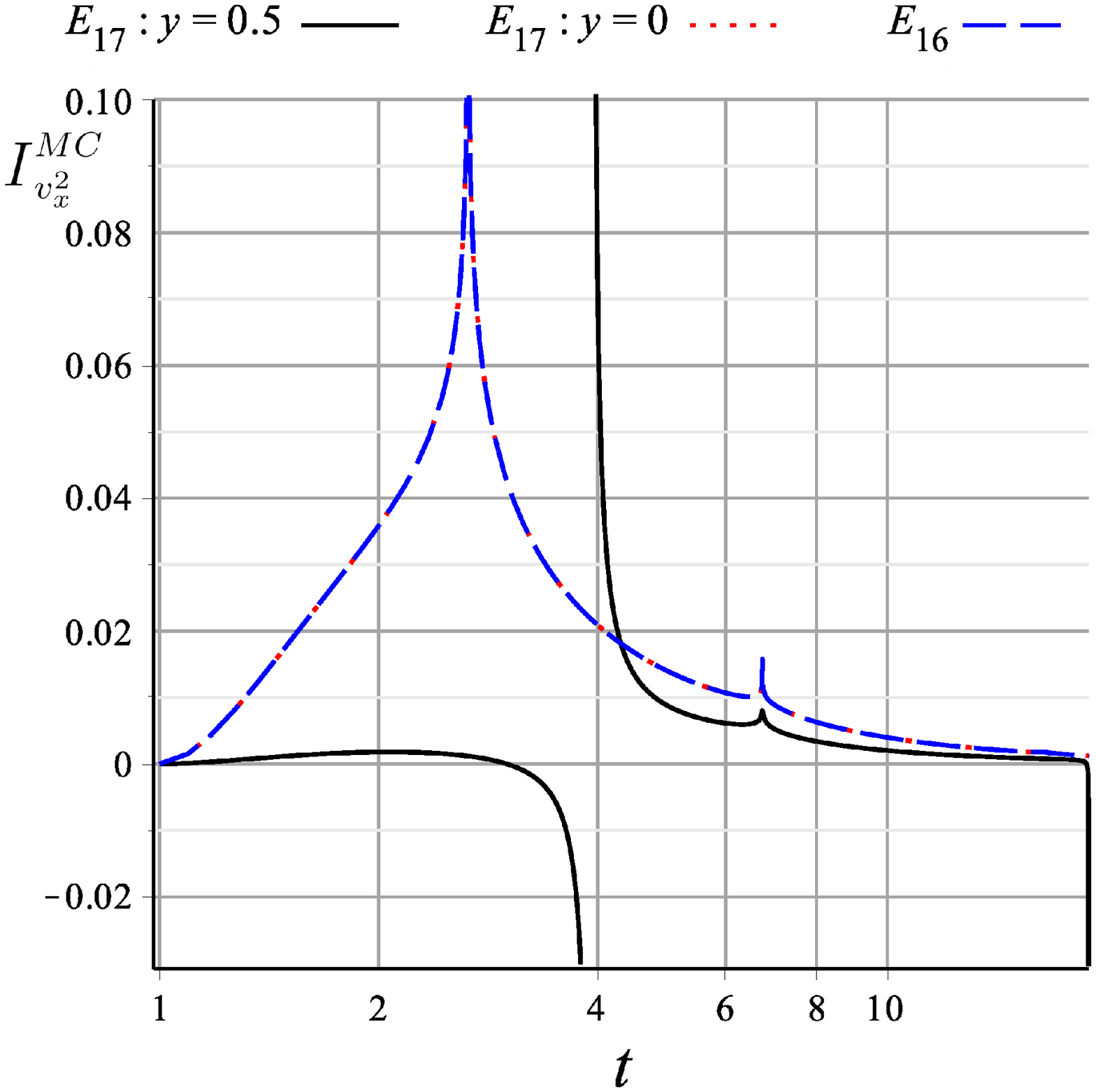}}    
    &    \resizebox{\halfsize}{!}{\includegraphics{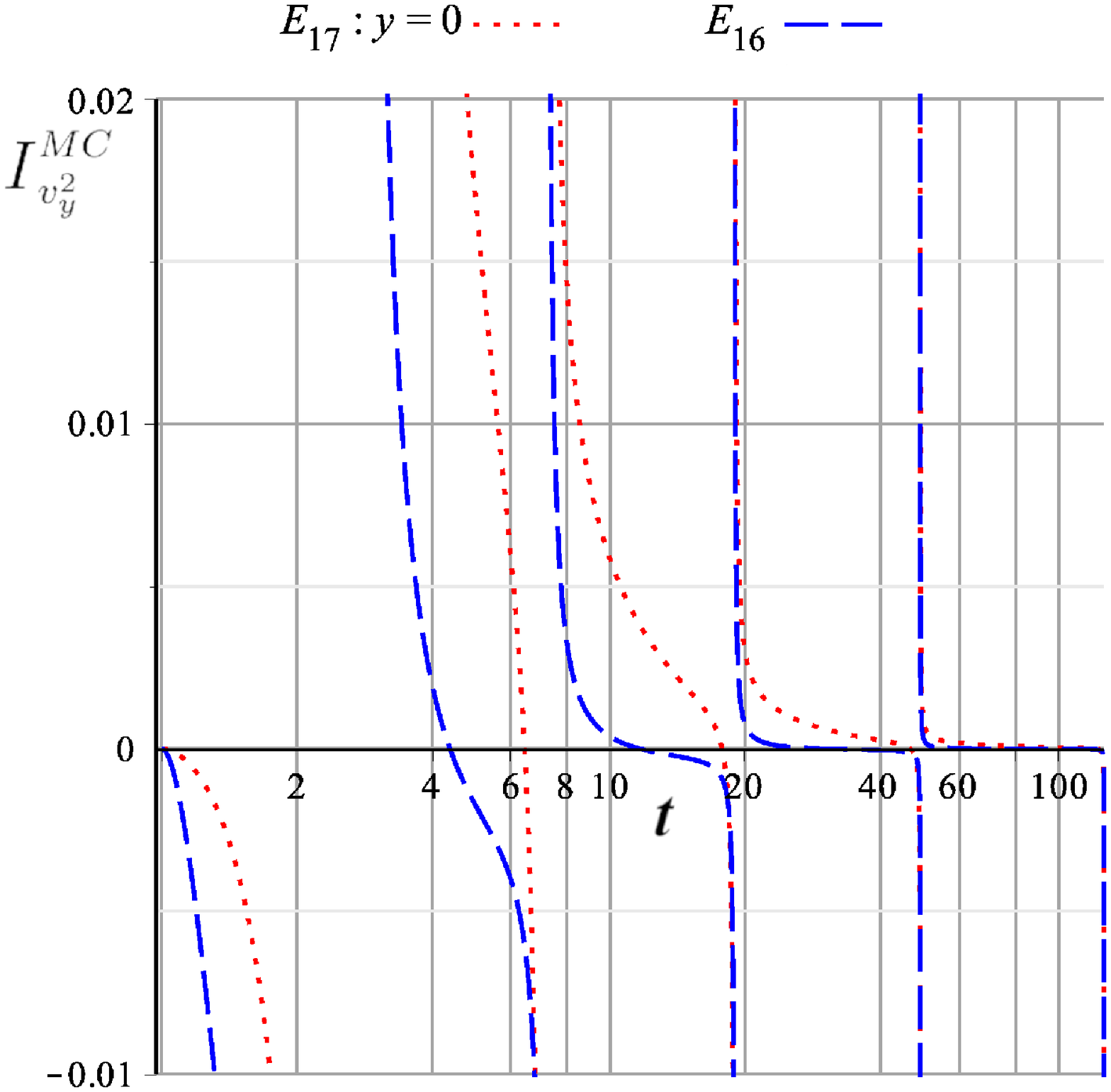}}  \\
       (a) & (b)
    \end{tabular}
    \end{center}
    \caption{Time evolution of orientability indicators  for the point charge.
    In (a) the curve for  indicator $\mbox{\large $I$}_{v^2_x}^{E_{17}}$ for $y=0$, shown as
    a dotted  line, coincides with the curve for  $\mbox{\large $I$}_{v^2_x}^{E_{16}}$,
    depicted as a  dashed line. For $y=1/2$, the solid curve for $\mbox{\large $I$}_{v^2_x}^{E_{17}}$
     is now different from the one for $\mbox{\large $I$}_{v^2_x}^{E_{16}}$.
     In (b) the orientability indicators $\mbox{\large $I$}_{v^2_y}^{E_{17}}$ and
     $ \mbox{\large $I$}_{v^2_y}^{E_{16}}$ are  different even for $y=0$.
\label{Fig1}}
\end{figure*}

Figure~\ref{Fig1} shows orientability indicators  as functions of cosmic time for the  point charge.
In panel (a) the time evolution of $ {\mbox{\large $I$}_{v^2_x}^{E_{16}}}$ given
by~\eqref{dispersion-orientable-x-E16} is shown as a dashed line and that of
${\mbox{\large $I$}_{v^2_x}^{E_{17}}}$, given by~\eqref{dispersion-x-E17}, is displayed as the
dotted line for $y=0$. These indicators coincide. In panel (a) we also show as the solid
line the  indicator $ {\mbox{\large $I$}_{v^2_x}^{E_{17}}}$ for $y=1/2$, exhibiting the
global inhomogeneity effect for the $E_{17}$ manifold. 
In panel (b) of Fig.~\ref{Fig1}  the component of  orientability indicator
$ {\mbox{\large $I$}_{v^2_y}^{E_{16}}}$ given by~\eqref{dispersion-orientable-yz-E16}
is represented by a dashed line, whereas the component ${\mbox{\large $I$}_{v^2_y}^{E_{17}}}$,
defined by Eq.~\eqref{dispersion-y-E17}, is depicted as a dotted line.
Now these indicator curves are distinct. 
Thus, it is possible to tell the two topologies apart by means of this component of the indicator.
Nevertheless, as will be seen  below, the dipole is  much more sensitive to non-orientability.
The behaviors of the components ${\mbox{\large $I$}_{v^2_z}^{E_{17}}}$
and~$ {\mbox{\large $I$}_{v^2_z}^{E_{16}}}$ are not shown because the corresponding curves
coincide  when $y=0$, as can be directly checked from equations~\eqref{dispersion-z-E17} and~\eqref{dispersion-orientable-yz-E16}, although they
are distinct if $y \neq 0$.

\begin{figure*}[h]
 \begin{center}
\begin{tabular}{c c}
        \resizebox{\halfsize}{!}{\includegraphics{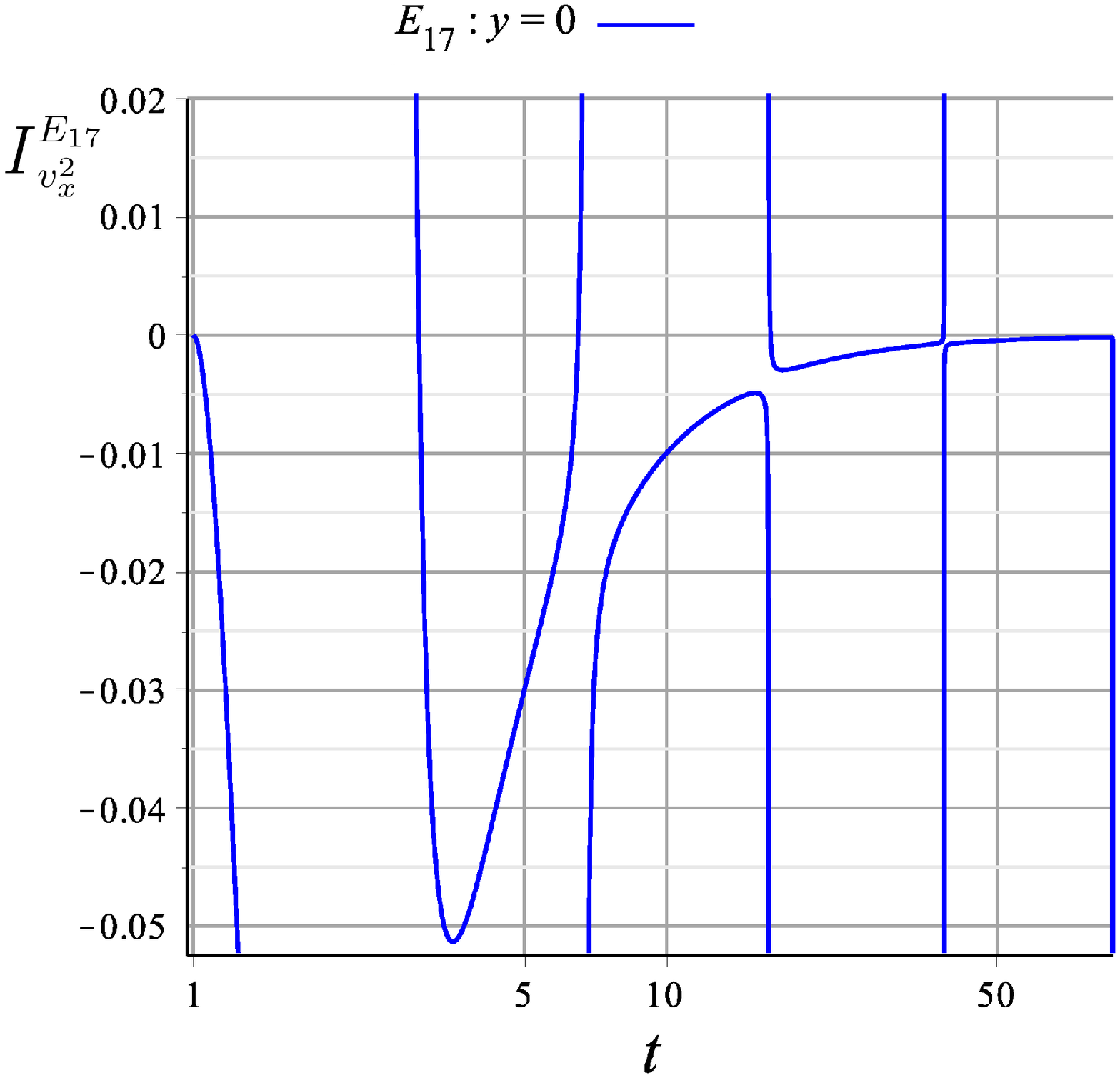}}    
    &    \resizebox{\halfsize}{!}{\includegraphics{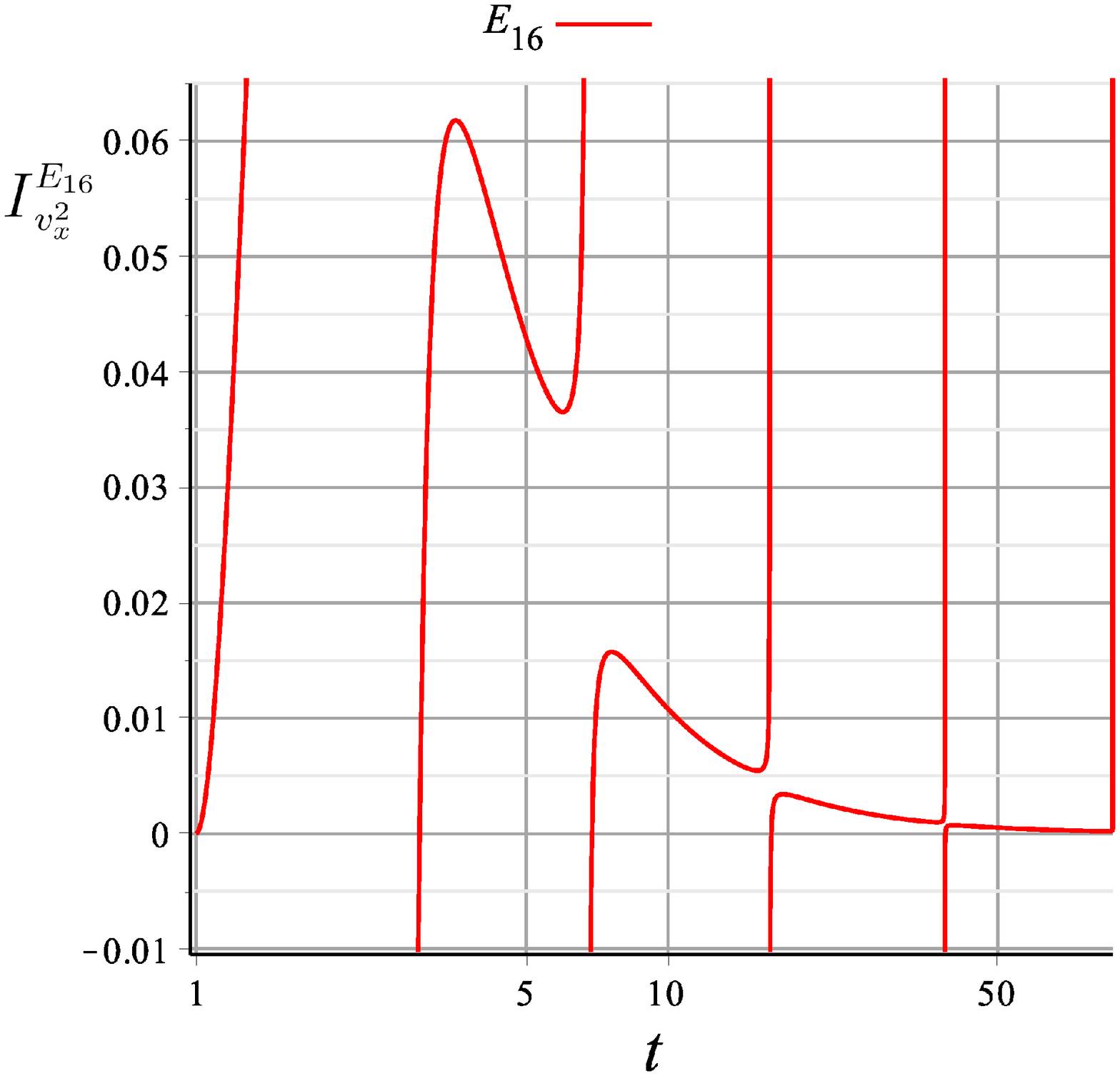}}  \\
       (a) & (b)
    \end{tabular}
    \end{center}
    \caption{Under the same conditions as in Fig.~\ref{Fig1}(b), but now for a point dipole,
    panel (a) shows the indicator $\mbox{\large $I$}_{v^2_x}^{E_{17}}$, whereas panel (b) displays
      $\mbox{\large $I$}_{v^2_x}^{E_{16}}$. We have intentionally made separate plots for the
      $E_{16}$ and  $E_{17}$ topologies to emphasize the repetitious inversion pattern
      roughly resembling $\cup$ followed by  $\cap$  in the case of the non-orientable
      spatial topology.
      The indicator  $\mbox{\large $I$}_{v^2_x}^{E_{16}}$ exhibited in panel (b) displays no inversion pattern.
\label{Figura2}}
\end{figure*}

\begin{figure*}[htpb]
 \begin{center}
\begin{tabular}{c c}
\resizebox{\halfsize}{!}{\includegraphics{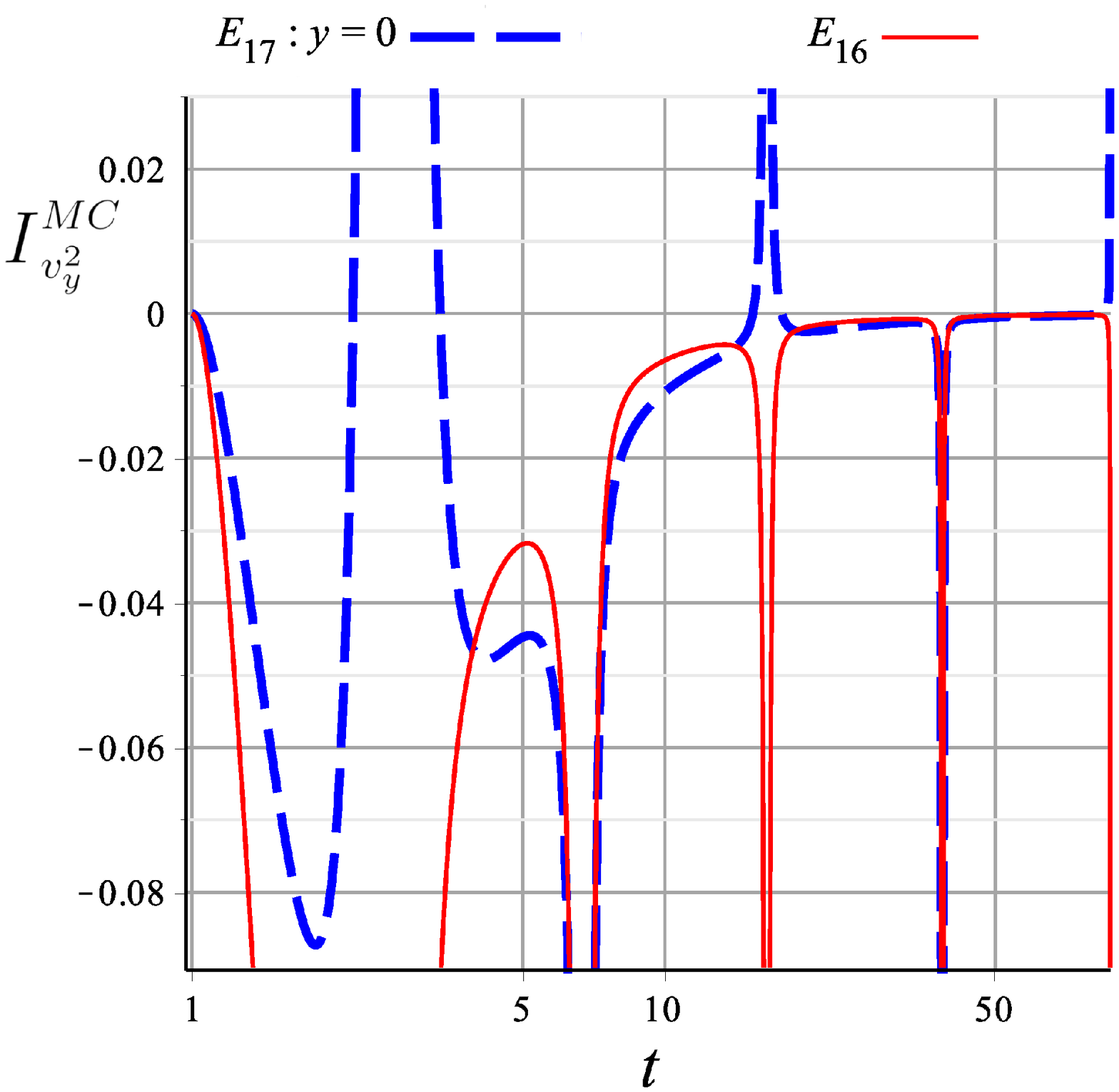}}    
    &    \resizebox{\halfsize}{!}{\includegraphics{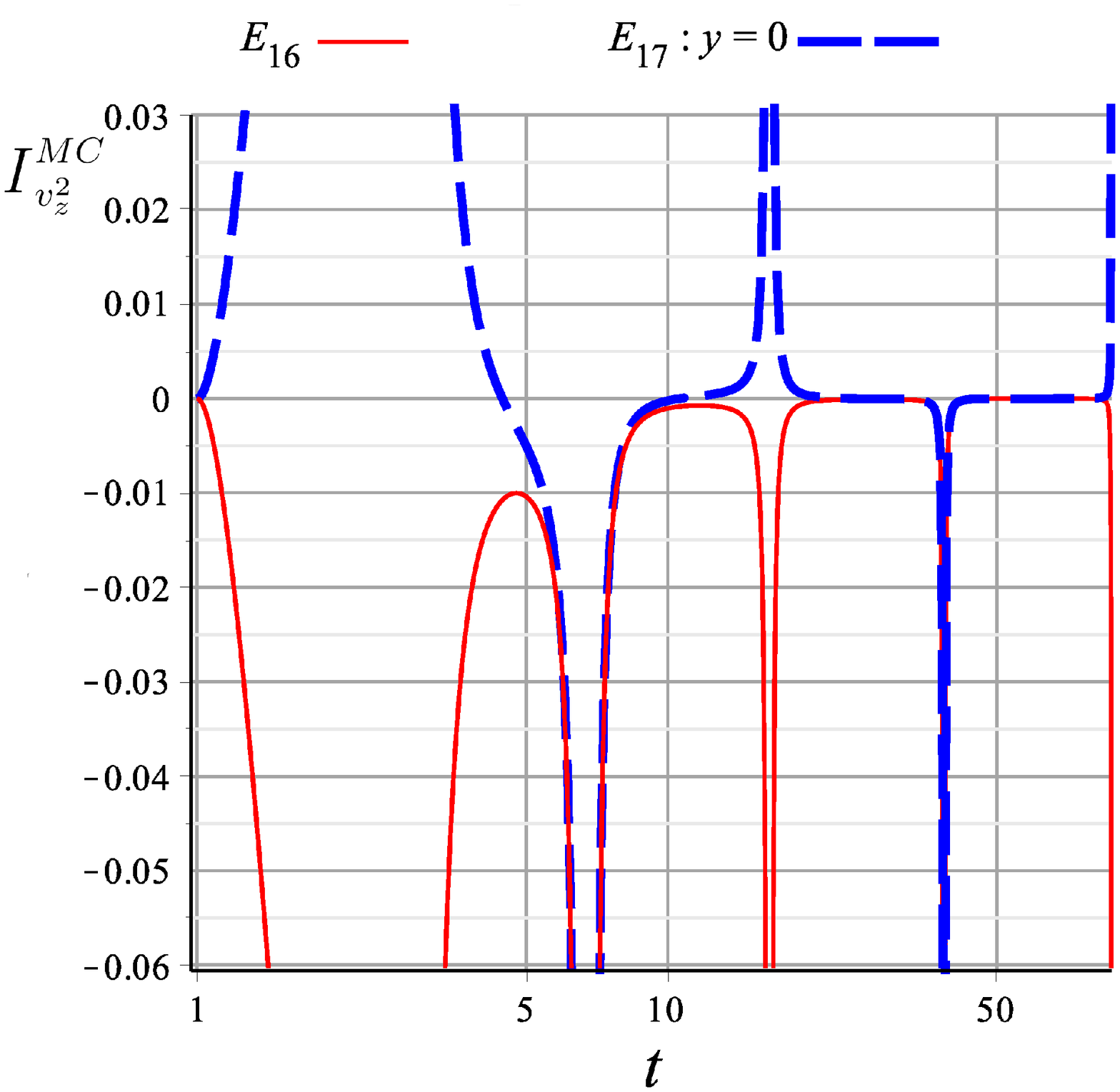}}  \\
       (a) & (b)
  \end{tabular}
    \end{center}
    \caption{Under the same conditions as in Fig.~\ref{Figura2}, and also for a point dipole,
    panel (a) shows  $\mbox{\large $I$}_{v^2_y}^{E_{17}}$  as a dashed line and
    $\mbox{\large $I$}_{v^2_y}^{E_{16}}$ as a solid line. Panel (b) displays
    $\mbox{\large $I$}_{v^2_z}^{E_{17}}$  as  a dashed line and
    $\mbox{\large $I$}_{v^2_z}^{E_{16}}$ as a  solid line. For FRW spacetime with the
    non-orientable spatial topology, $E_{17}$,  there is an inversion pattern resembling
    successive upward and downward  ``horns". Similar alternating horn-like inversion
    patterns also arise in static Minkowski spacetime with  $E_{17}$
    topology~\cite{Lemos:2020ogj}, but here their shape is modified by the dynamical scale factor.
\label{Fig3}}
\end{figure*}

\subsection{Non-orientability: point dipole case}

Figure~\ref{Figura2} shows the time evolution of orientability indicators for the point dipole.
Fig.\ref{Figura2}(a) displays the dipole indicator $ \mbox{\large $I$}_{v^2_x}^{E_{17}}$,
 given by Eq.~\eqref{eqdispersion-dipole-yx-final-E17}, and  Fig.\ref{Figura2}(b) shows
 $ \mbox{\large $I$}_{v^2_x}^{E_{16}}$, obtained from equation~\eqref{eqdispersion-dipole-yx-final-E16}.
 We have  intentionally exhibited in separate plots the indicators for  the  manifolds $E_{17}$
 and $E_{16}$ in order to highlight the repetitious pattern  roughly resembling  $\cup$ followed
 by  $\cap$ in  the non-orientable case.
 Similar repetitious  inversion  patterns are also present in Minkowski spacetime with
 $E_{17}$ topology~\cite{Lemos:2020ogj}. Here the shape of the curves is
 modified by the dynamical scale factor $a(t)$, though.
 Just as in the Minkowski case, the orientability indicator is already sensitive to
 non-orientability even for $y=0$. The inversion pattern exhibited by $E_{17}$ is
 qualitatively different from the pattern for $E_{16}$, which is not characterized by
 successive inversions, making it possible to identify  the non-orientable case in itself.

Figure~\ref{Fig3} exhibits the two remaining components of the  dipole indicators.
Fig.~\ref{Fig3}(a) displays as a dashed line the dipole  indicator
$\mbox{\large $I$}_{v^2_y}^{E_{17}}$ given by~\eqref{eqdispersion-dipole-yy-final-E17},
together with  $ \mbox{\large $I$}_{v^2_y}^{E_{16}}$, defined
by~\eqref{eqdispersion-dipole-yy-final-E16}, which is depicted as a solid line.
Panel (b) of Fig.~\ref{Fig3} shows as a dashed line the orientability indicator
$\mbox{\large $I$}_{v^2_z}^{E_{17}}$ given by~\eqref{eqdispersion-dipole-yz-final-E17},
as well as  $\mbox{\large $I$}_{v^2_z}^{E_{16}}$, defined
by~\eqref{eqdispersion-dipole-yz-final-E16}, as a dotted line.
Both panels reveal the same inversion pattern roughly resembling successive
upward and downward  ``horns".
Similar alternating horn-like inversion patterns emerge in static Minkowski
spacetime with $E_{17}$ topology~\cite{Lemos:2020ogj}. Now, however, the shape
of the curves is modified by the dynamical scale factor $a(t)$.
This distinctive pattern allows one to recognize the non-orientability of
$3-$space per se.

\subsection{Non-orientabily: summary of findings}
Stochastic motions of a point electric charge under quantum electromagnetic fluctuations
give rise to the orientability indicators defined by equations \eqref{dispersion-x-E17} to
\eqref{dispersion-orientable-yz-E16}. Fig. \ref{Fig1}(b) shows that the $y$-component of
the  orientability indicator for $E_{17}$ is different from the one for $E_{16}$.
Therefore, one can distinguish one  topology from the other. But the curve patterns
for both topologies are qualitatively the same. This means that the identification of
a putative non-orientable topology requires a quantitative  comparison of its
evolution curves with those for the counterpart orientable topology.
In short, we are unable to spot a non-orientable topology \textsl{per se}
by means of the stochastic motions of a point charged particle.

Things become more appealing when one considers  the stochastic motions
of a point electric dipole, whose corresponding non-orientability
indicators are given by equations~\eqref{eqdispersion-dipole-yx-final-E17}
to~\eqref{eqdispersion-dipole-yz-final-E16}. A comparison of the
$x$-component of the orientability indicators, exhibited in Fig.~\ref{Figura2},
already shows  a repetitious inversion pattern of roughly resembling
$\cup$ followed by  $\cap$
for $E_{17}$ that distinguishes it from $E_{16}$. This distinction is much
more pronounced when the remaining components of the indicators for
$E_{17}$ and $E_{16}$ are compared, as in Fig.~\ref{Fig3}.
The conspicuous  pattern roughly
resembling  alternating upward and downward ``horns"
for $E_{17}$, which is absent from the curves for $E_{16}$, enables
us to identify the non-orientability of $E_{17}$ by itself, without
the need to compare its  indicator curves with those for its orientable
counterpart. This is the greatest advantage of the dipole over the
point charge in the search for detecting non-orientability through
 stochastic motions of point-like objects under quantum vacuum
fluctuations of the electromagnetic field.

Contrarily to the dipole case, the point-charge case does
not exhibit such prominent qualitative features and hence appears to be
of hardly any practical use.

\section{Closing remarks and conclusions} \label{Finals}
In the framework of general relativity
the Universe is modelled as a four-dimensional differentiable manifold
$\mathcal{M}_4$ endowed with the FRW metric~\eqref{RWmetric} that expresses
geometrically two basic assumptions of the cosmological modeling, namely the
existence of a cosmic time $t$, which emerges from Weyl's principle, and the
cosmological principle, which in turn ensures that the $3-$dimensional space $M_3$
is geometrically homogeneous and isotropic.
The FRW metric does not specify the topology of the underlying spacetime manifold
$\mathcal{M}_4$ or of the corresponding spatial ($t= \mbox{const}$) sections $M_3$,
which can in principle be found through observations.
So far, however, direct searches for a nontrivial topology of $M_3$ using CMB data from
WMAP and Planck have found no convincing evidence of multiple connectedness 
below the radius of the last scattering surface~\cite{2014,2016,%
Cornish:2003db,ShapiroKey:2006hm, Bielewicz:2010bh,Vaudrevange:2012da,Aurich:2013fwa}.%
\footnote{This does not exclude the possibility of a FRW universe with
a detectable nontrivial cosmic topology~\cite{Gomero:2016lzd,%
Bernui:2018wef,Aurich:2021ofm}. }
In this work, rather than focusing on determining the topology of the
spatial sections $M_3$ of FRW spacetime, we have investigated its global
property of orientability.   

In the physics at daily and even astrophysical length and time scales we do
not find any sign or hint of non-orientability of  $3-$space.
At the cosmological scale, in order to disclose  spatial  
non-orientability,  global trips around the whole $3-$space would be needed
to check for orientation-reversing closed paths.  
Since such  global journeys across the Universe are not feasible one might
think that spatial orientability cannot be probed. 
We note, however, that the determination  of the spatial
topology through, for example, the so-called circles-in-the-sky,
would bring out as a bonus an answer as to  $3-$space orientability
at the cosmological scale.

On the other hand, at a theoretical level of reasoning, it is often
assumed that the spacetime manifold is separately time and space orientable.
As we have mentioned, the arguments supporting orientability combine the
space-and-time universality of the local  physical  laws\footnote{We note that space universality can be looked upon as a
topological assumption of global homogeneity of $M_3$. So, all elements
of the covering group $\Gamma$  are translations, and therefore spatial
universality by itself rules out non-orientable $3-$spaces.}
with a well-defined local arrow of time along with some local results 
from discrete symmetries in particle physics~\cite{Hawking:1973uf,1979grec.conf..212G}.
Of course one is free to resort to such  reasonings, but it is reasonable
to expect that the ultimate answer to questions
regarding the orientability of spacetime should rely on cosmological
observations or local experiments, or might come from a fundamental
theory of physics.

In this paper we have investigated whether electromagnetic quantum vacuum
fluctuations can be used to access the spatial orientability of a FRW expanding
spacetime, extending therefore the results of the recent paper~\cite{Lemos:2020ogj},
where the question of spatial orientability of Minkowski (static) spacetime was
examined. 
To this end, we have studied the stochastic motions of point-like objects
under quantum electromagnetic fluctuations in FRW flat spacetime with the
orientable $E_{16}$ (slab)  and non-orientable $E_{17}$ (slab with flip)
space topologies (cf.\ Tables~\ref{Tb-2-Orient_and_Non_orient}
and~\ref{Tb-Spatial-separation}).

The statistical indicator $\mbox{\large $I$}_{v^2_i}^{MC}$ [Eq.~\eqref{new-ind}], which
measures the departure  of the mean square velocity dispersion for  point-like
objects in the multiply-connected topology from its value for the
simply-connected covering space, has been shown to be suitable to reveal
spatial orientability of flat FRW spacetime.
In the case of a charged particle, we  have derived  expressions
\eqref{dispersion-x-E17}~--~\eqref{dispersion-orientable-yz-E16}
for the  orientability indicator $\mbox{\large $I$}_{v^2_i}^{MC}$
 for the 
$E_{17}$ and  $E_{16}$ space topologies. 
Similarly, we have derived  expressions~\eqref{eqdispersion-dipole-yx-final-E17}
~--~\eqref{r-rx-ry-coincidence-E17} for the indicator~\eqref{new-ind} corresponding to
a point electric dipole
oriented in the flip  direction of  $E_{17}$ 
topology, and also  expressions~\eqref{eqdispersion-dipole-yx-final-E16}~--%
~\eqref{eqdispersion-dipole-yz-final-E16} for the dipole in $3-$space
with the orientable $E_{16}$ topology.

The expressions for the orientability indicators for the particle
and the dipole in  $E_{17}$  and  $E_{16}$ spatial topologies
hold for an arbitrary scale factor $a(t)$, which is determined by
the  gravitational theory.  
In this way, to concretely study the time evolution of the orientability
indicator  $\mbox{\large $I$}_{v^2_i}^{MC}$
we have assumed in Section~\ref{casestudy} the general relativity theory
and also that the matter content consists of a
single-component barotropic perfect fluid  with equation of state $p = w \rho$,
with the equation of state parameter $w$ such that $| w | < 1$.
Under these assumptions we have made  figures~\ref{Fig1} to~\ref{Fig3}.

Figure~\ref{Fig1}(a) illustrates the  topological inhomogeneity effect
of $E_{17}$ topology and makes it clear that $y=0$ is the appropriate
particle's position in $E_{17}$ for comparison of the evolution of
the orientability indicator in $E_{16}$ and $E_{17}$ topologies.
Figure~\ref{Fig1}(b) shows that  it is possible to distinguish
the orientable from the non-orientable topology by comparing the time
evolution of the respective $y-$components of the orientability indicator:
they give rise to different  evolution curves for distinct topologies.

A  more ambitious goal is that of finding a way to decide about
the orientability of a given spatial manifold in itself, without
having to make a comparison of the results for a non-orientable
space with those for its orientable counterpart.
We have addressed this matter and have shown that the
stochastic motions of a point electric dipole can be used
to disclose the putative non-orientability of a generic
$3-$space per se.
To this end, under the premises put forward in Section~\ref{casestudy}
(flat spacetime in general relativity with perfect fluid source)
we have used expressions
\eqref{eqdispersion-dipole-yx}~--~\eqref{r-rx-ry-coincidence-E17} and
\eqref{eqdispersion-dipole-yx-final-E16}~--~\eqref{eqdispersion-dipole-yz-final-E16}
for the stochastic motions of the dipole in FRW spacetime with, respectively,
the non-orientable $E_{17}$ and orientable $E_{16}$ spatial topologies to
plot Figures~\ref{Figura2} and \ref{Fig3}. These figures show
that an inversion pattern for the  orientability  
indicator curves  comes about in the case of the non-orientable $E_{17}$ topology,
implying that the non-orientability of $E_{17}$ can be detected per se.%
\footnote{The curves for $E_{16}$ display a repetition pattern, which, however
is not an inversion pattern.}

Our results for the non-orientability indicator hold for any scale factor
$a(t)$. For the case studied --- flat FRW geometry in the context of general
relativity;  matter content described by a perfect fluid with equation of state $p = w\rho$ with
 $w = -0.299$ fixed from the Planck satellite observational data --- a  non-orientability signature in the form of an inversion pattern
was found.
All we can be sure of is that there is an inversion pattern
in this particular case. In the framework of some other metric
gravitational theory or for other scale factors one has to re-examine the
question in order to find out whether the scalar factor  would preserve,
modify or even destroy the inversion pattern.

An expected result from the beginning of this work was that the role played
by the topology on the stochastic motions of particles would depend crucially
on the topological compact length $L$, which in turn gives rise 
to a lower bound for the time scale required to test orientability. 
However, the time scale involved in  Figs. \eqref{Fig1}~--~\eqref{Fig3} makes it clear that
to access the inversion patterns of the orientability indicators one needs
a relatively long period of time, typically the time needed
 to travel across quite a few  $L$s. A small topological length scale 
is  expected, for example, in the primordial universe. 
An open  question is whether those velocity fluctuations would leave
traces that could be extracted from today's observational data, as for example
from CMB maps, making it potentially possible to unveil information on
 $3-$space orientability. This is a nontrivial and important issue
beyond the scope of the present paper, though.

\begin{acknowledgements}
M.J. Rebou\c{c}as acknowledges the support of FAPERJ under a CNE E-26/202.864/2017 grant,
and thanks CNPq for the grant under which this work was carried out.
We thank  C.H.G.  Bessa for fruitful discussions.
M.J.R. is also grateful to A.F.F. Teixeira for interesting comments, and also for reading
the manuscript and indicating typos.
\end{acknowledgements}


\end{document}